\documentclass[journal=jacsat,manuscript=article]{achemso}
\setkeys{acs}{articletitle = true}

\pdfoutput=1
\usepackage{array}
\usepackage{booktabs}
\usepackage[font=small,labelfont=bf]{caption}
\usepackage[version=3]{mhchem} 
\usepackage{hyperref}
\usepackage{graphicx}
 \graphicspath{ {./} }
\usepackage{epstopdf}
\usepackage[compatibility=false]{caption}
\DeclareCaptionFont{quackfont}{\fontfamily{ptm}\fontsize{9pt}{9pt}\selectfont}
\usepackage[font=quackfont]{subcaption}
\usepackage{cleveref}
\author{Aditya N. Singh}
\author{Arun Yethiraj}
\email{yethiraj@wisc.edu}
\affiliation[UW-Madison]
{Theoretical Chemistry Institute and Department of Chemistry, 1101 University Avenue, University of Wisconsin-Madison, Madison, Wisconsin 53703}

\title[An \textsf{achemso} demo]
  {The Driving Force for the Complexation of Charged Polypeptides}

\abbreviations{IR,NMR,UV}

\begin{document}
\begin{tocentry}
\includegraphics[trim={5cm 1.8cm 9.7cm 4.6cm},clip,width=\linewidth]{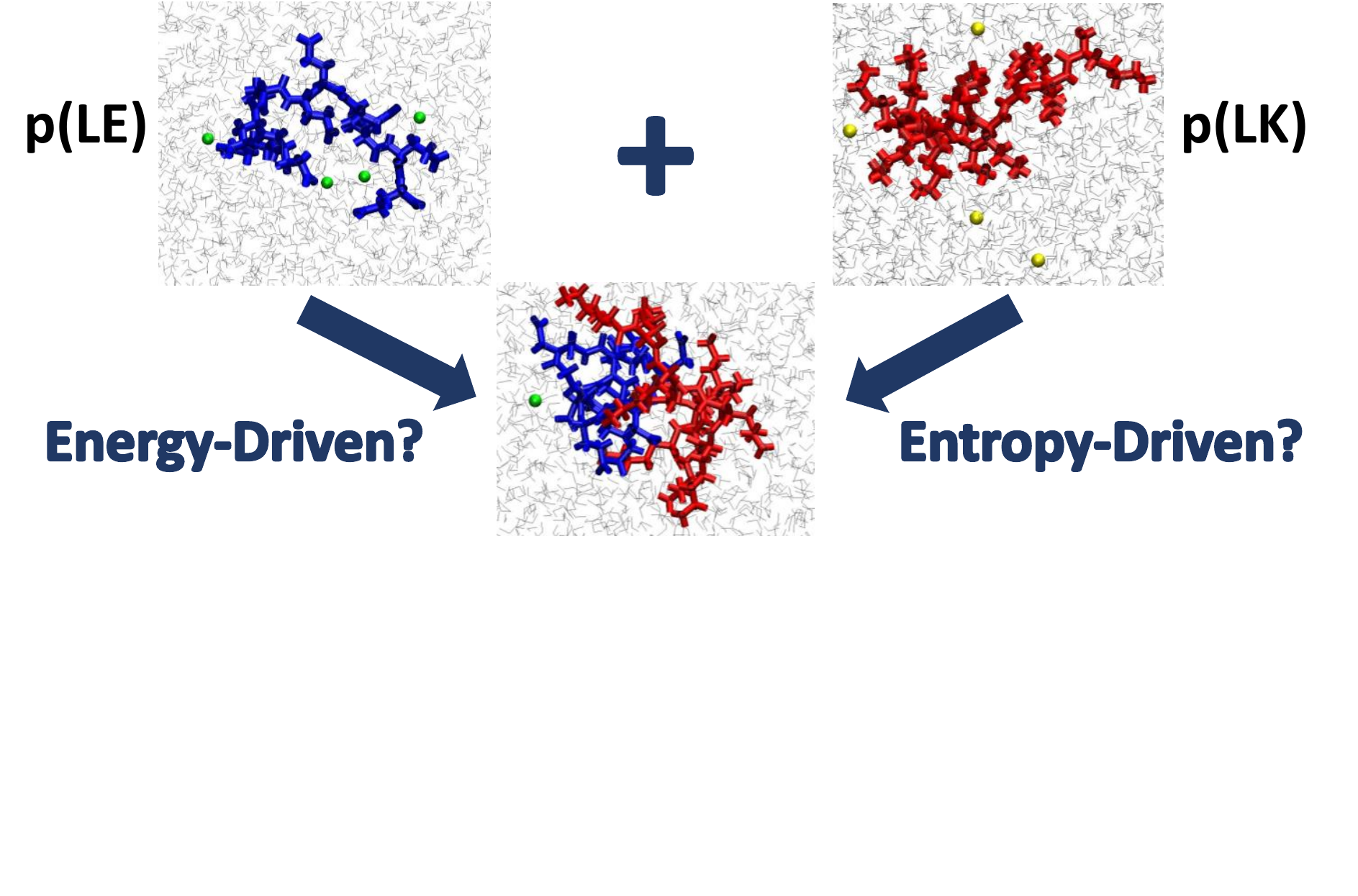}

\end{tocentry}
\begin{abstract}
The phase separation of oppositely-charged polyelectrolytes in solution is of current interest .  In this work we study the driving force for polyelectrolyte complexation using molecular dynamics simulations.  We calculate the potential of mean force between poly(lysine) and poly(glutamate) oligomers using three different forcefields, an atomistic force field and two coarse-grained force fields.  There is qualitative agreement between all forcefields, i.e., the sign and magnitude of the free energy and the nature of the driving force are similar, which suggests that the molecular nature of water does not play a significant role.  For fully charged peptides,
we find that the driving force for association is entropic in all cases when small ions either neutralize the poly-ions, or are in excess.
The removal of all counterions switches the driving force, making complexation energetic.  This suggests that the entropy of complexation is dominated by the counterions.  When only 6 residues of a 11-mer are charged, however, the driving force is enthalpic in salt-free conditions.  The simulations shed insight into the mechanism of complex coacervation and the importance of realistic models for the polyions.
\end{abstract}

\section{Introduction}

Oppositely-charged polyelectrolytes in aqueous solution can undergo a liquid-liquid phase separation to form a polymer-rich and a polymer-poor phase, through a process called complex coacervation.\cite{decher1997fuzzy,van2011polyelectrolyte}.   The polymer-rich ``coacervate" phase can be formed using DNA, polypeptides or polymers in microemulsions\cite{C4SM02336F}, and the concentration of polyelectrolytes in the coacervate can be changed by tuning the pH\cite{doi:10.1021/bm300835r}, charge density\cite{doi:10.1002/jcp.1030490404} and chirality\cite{C4SM02336F} of the polyelectrolytes, ionic strength\cite{PhysRevLett.105.208301}, temperature\cite{doi:10.1021/ma902144k} and the concentration of salt.\cite{doi:10.1002/jcp.1030490404}.   Complex coacervates have a wide range of applications from microencapsulation of food products\cite{SILVA2014} to drug-delivery\cite{doi:10.1517/17425247.2014.941355}, protein purification\cite{Wang1996356} and dispersion of cells\cite{KimE847}. 

The thermodynamics of this phenomenon is not completely understood.\cite{van2011polyelectrolyte,rathee2018role,doi:10.1021/jacs.5b11878} 
One expects a pairing of oppositely charged polyions in the dilute and co-acervate phase, given the strong electrostatic attraction.  
Multiple experimental\cite{DEKRUIF2004340,gummel2007,PRIFTIS201339,doi:10.1021/la100705d} and computational studies\cite{doi:10.1063/1.2178803,rathee2018role,doi:10.1021/bm201113y} argue that complex coacervation has a strong favorable entropic contribution, which could arise from the gain in translational entropy of the counterions when the polyions are complexed.

Several theoretical models have been devised to understand complex coacervation, including the Voorn-Overbeek (VO) theory\cite{doi:10.1002/jcp.1030490404,michaeli1957phase}, approaches based on the random phase approximation (RPA)\cite{doi:10.1021/ma00217a015}, field theoretic methods\cite{doi:10.1002/polb.21334, doi:10.1063/1.2936834, doi:10.1021/acs.macromol.6b02160} and integral equation theories\cite{doi:10.1021/acs.macromol.5b01027}.  
The theoretical methods rely on simple models and on approximations and molecular dynamics simulations are restricted by the possible lengthscales and timescales accessible\cite{SING20172}.  A molecular dynamics simulation\cite{C4SM02336F} of two polyions using the 
CHARMM22\cite{doi:10.1002/jcc.21287} forcefield does show complex coacervation, but does not establish the driving force.

Two important concepts in polyelectrolyte complexation are counterion condensation and counterion release.  The original Manning formulation 
\cite{doi:10.1063/1.1672157,manning2} of counterion condensation considered a single infinite line of charges and neutralizing counterions.  An important parameter is the ratio $\xi = l_B/l$ where $l$ is the spacing between charges and $l_B$ is the Bjerrum length, defined as $l_B = e^2 / 4 \pi \epsilon_0 \epsilon k_B T$ where $e$ is the charge of an electron, $\epsilon_0$ is the permittivity of free space, $\epsilon$ is the dielectic constant,  $k_B$ is Boltzmann's constant and $T$ is the temperature.  For an infinitely thin and long line of charges, Manning found a divergence in the free energy for $\xi=1$ and postulated that counterions would ``condense" onto the polyion for $\xi > 1$, thereby partially neutralizing the polyion charge until $\xi < 1$.  The condensed counterions were assumed to be bound, and did not contribute to the translational entropy 
\cite{doi:10.1063/1.1672157} or the self-diffusion constant \cite{manning2} of the counterions.  

For more realistic models, with polyions of finite length or excluded volume, there is no divergence in the free energy as the linear charge density increases.  The electrostatic attraction does result in a higher concentration of counterions near the polyion compared to the bulk.  The term condensation is often used in this case although the counterions are not bound, and in fact have a very short lifetime (in this work we find it is of the order of picoseconds) near the polyion.  The counterion self-diffusion constant is not zero and, since all counterions sample the entire volume, there is no decrease in {\em translational} entropy.

Counterion release is often cited as a mechanism for entropy gain due to complexation \cite{doi:10.1063/1.2178803}.  The picture is that the counterions are condensed on the polyions when they are isolated.   When they complex, the electrostatic attraction between polyions is energetically favorable and the pairing of the polyions causes an essentially electro-neutral complex that does not attract either of the counterion species. As a consquence the concentration of counterions near the polyions is significantly reduced compared to the isolated polyions.  This is viewed as a release of counterions and often interpreted as resulting in an increase in the translational entropy of the counterions.  However, for the reasons mentioned above, this simple picture cannot be the complete story.  The two driving forces that are often discussed are the energetic attraction between polyions and the entropy gain due to the counterions\cite{rathee2018role,doi:10.1002/jcc.21287,doi:10.1063/1.2178803}.

The driving force for polyion complexation has been investigated in computer simulations by Ou and Muthukumar (OM) \cite{doi:10.1063/1.2178803}.
The studied uniformly charged polyions in a continuum solvent using Langevin dynamics simulations.  Starting with the two polyions far apart, they allowed them to complex and calculated the change in energy, which they identified as the internal energy change of the process.  The Hemholtz free energy change was then calculated via thermodynamic integration and the entropy change from the difference between the Helmholtz free energy and the internal energy.  The main conclusion of this work was that the driving force was energetic for weakly charged polyions ($\xi <1$)
but become entropic for strongly charged polyions ($\xi > 1$).  Similar results have been reported by Rathee et al.\cite{rathee2018role} for associatively charged polyions, where the charge state is determined by conditions of chemical equilbrium.

In this work, we investigate a system of two oppositely charged polypeptides with 10 residues each, and calculate the potential of mean force as a function of separation.  By performing the calculation at two temperatures we are able to obtain the entropic and energetic contributions to the free energy.  We investigate three different forcefields and compare their predictions. Two of the forcefields chosen for this study are coarse-grained, and the third is atomistic, and all of them have been used to study polyelectrolytes.\cite{doi:10.1021/bm500658w, TARAKANOVA2019100016, marrink2003mechanism, kim2013effect, doi:10.1021/bm201113y,C4CP04967E}.  We find that all the forcefields are in qualitative agreement with each other, i.e., the magnitude of the free energy and the nature of the driving force are similar,  suggesting that the model of water does not play a role.   When every residue of the peptides is charged the driving force is entropic in all cases except when no small ions are present.  When only 6 of the residues are charged, however, the driving force becomes energetic without added salt.  This is in qualitative agreement with the OM study \cite{doi:10.1063/1.2178803} in the sense that the driving force changes from energetic to entropic as the charge density is increased.  There are, however, some significant differences.  We can estimate $l$ by dividing the root-mean-square-end-to-end distance of the peptides by the number of charges.  For the atomistic model where every residue is charged this gives a value of $l$=0.225nm for polyglutamate and $l$=0.214nm for polylysine.  Since the Bjerrum length for the water model is 0.69nm this gives $\xi \approx$3.07 and 3.22 for the two cases.  For the case where the peptides have 11 residues, and 6 of the residues are charged, $\xi \approx 1.8$ and 2.2 for the two peptides, which is above the counterion condensation threshold.  Furthermore we see no evidence for counterion localization, in that all counterions have a non-zero self-diffusion coefficient and short lifetimes near the peptides.  

\section{Computational Methods}

\begin{figure}[h!]
  \centering
  \begin{subfigure}[b]{0.8\linewidth}
  \centering
    \includegraphics[trim={0cm 2cm 1cm 2cm},clip,width=.8\linewidth]{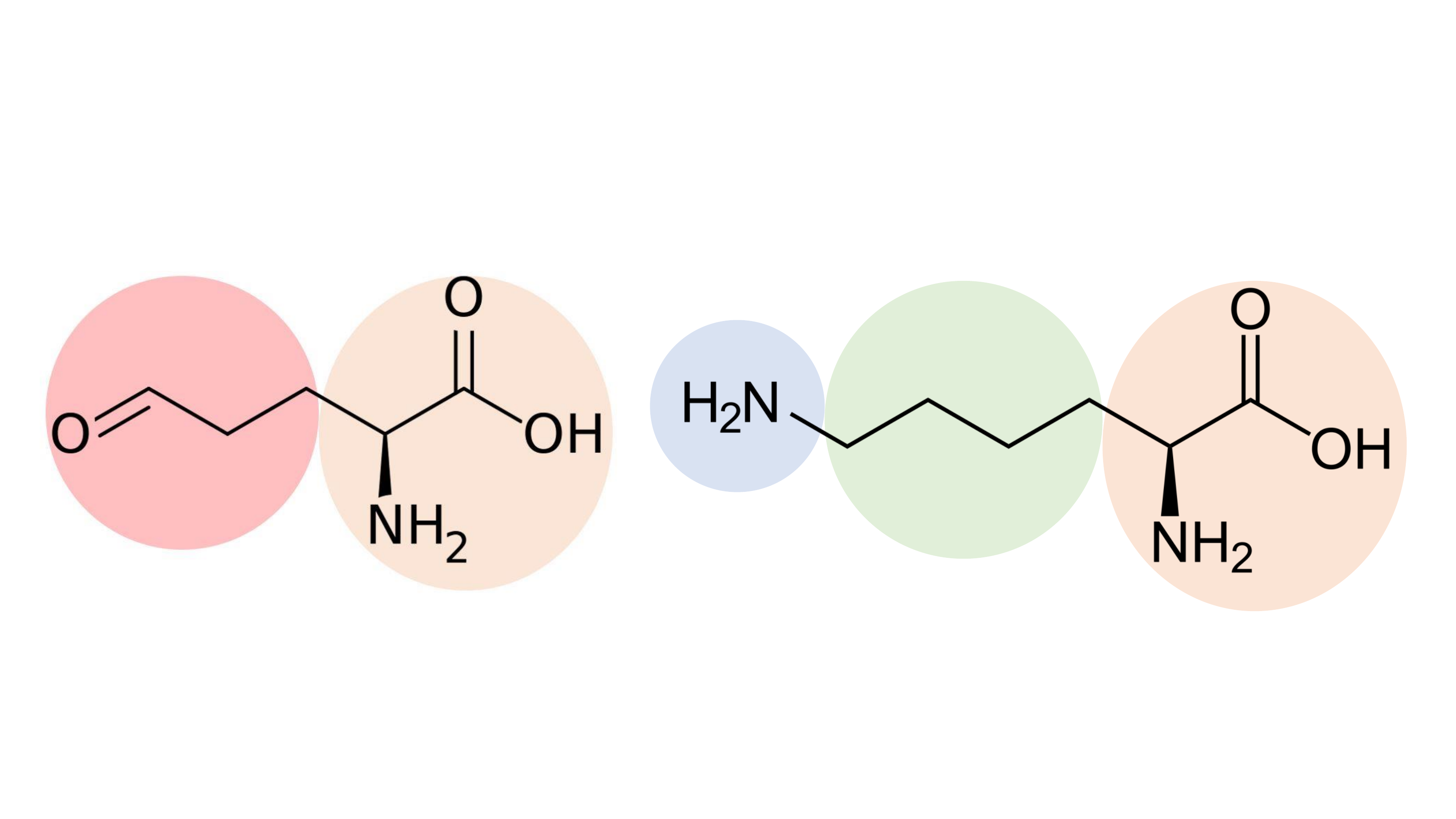}
    \caption{Coarse-graining used in the MARTINI forcefield for glutamic acid (left) and lysine (right). The orange beads correspond to the backbone and the blue, green, and red beads correspond to different side chains.}
  \end{subfigure}
  \begin{subfigure}[b]{0.4\linewidth}
    \includegraphics[width=\linewidth]{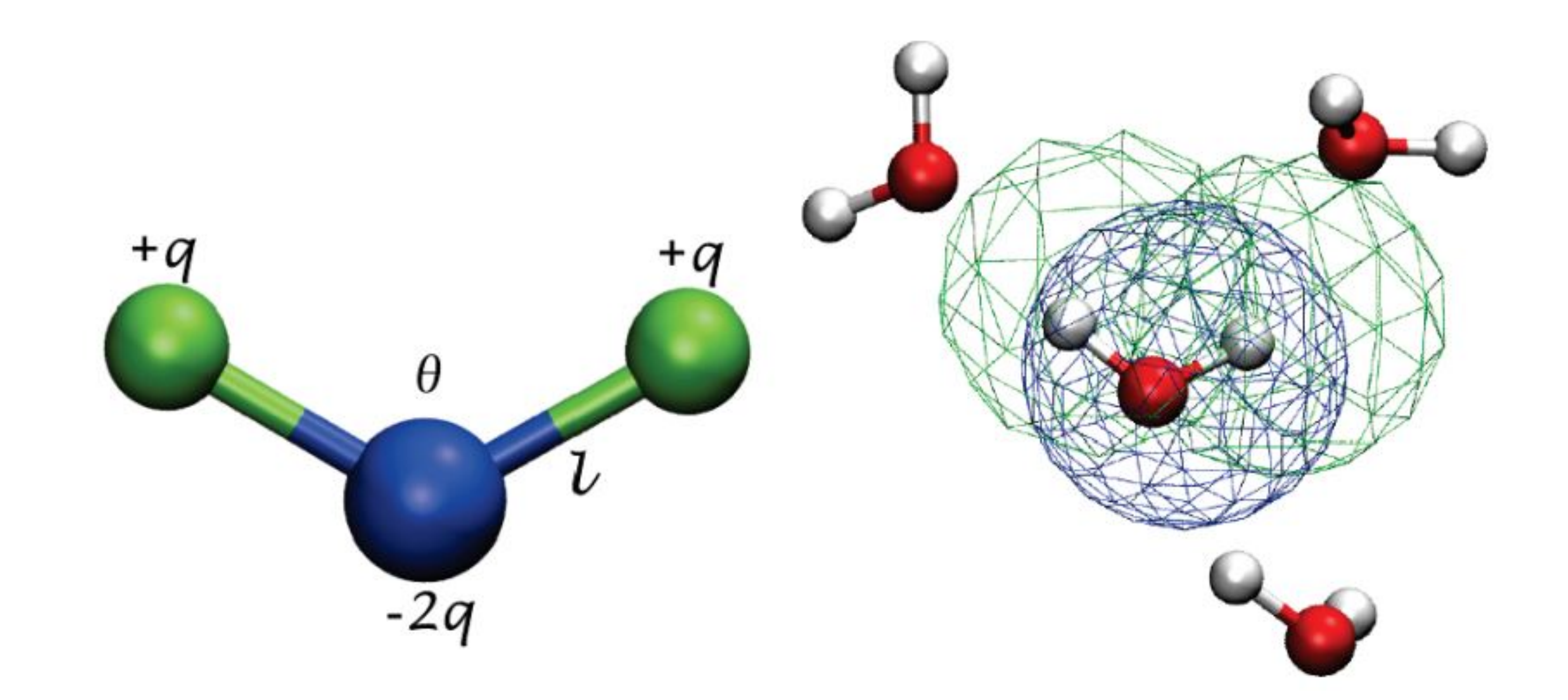}
    \caption{Coarse-graining of the Big Multipole Water model. 4 water clusters (right) are described using the water model (left). Reproduced with permission from the Journal of Physical Chemistry B.\cite{doi:10.1021/jp1019763}}
  \end{subfigure}
  \begin{subfigure}[b]{0.4\linewidth}
    \includegraphics[width=\linewidth,height=\textheight,keepaspectratio]{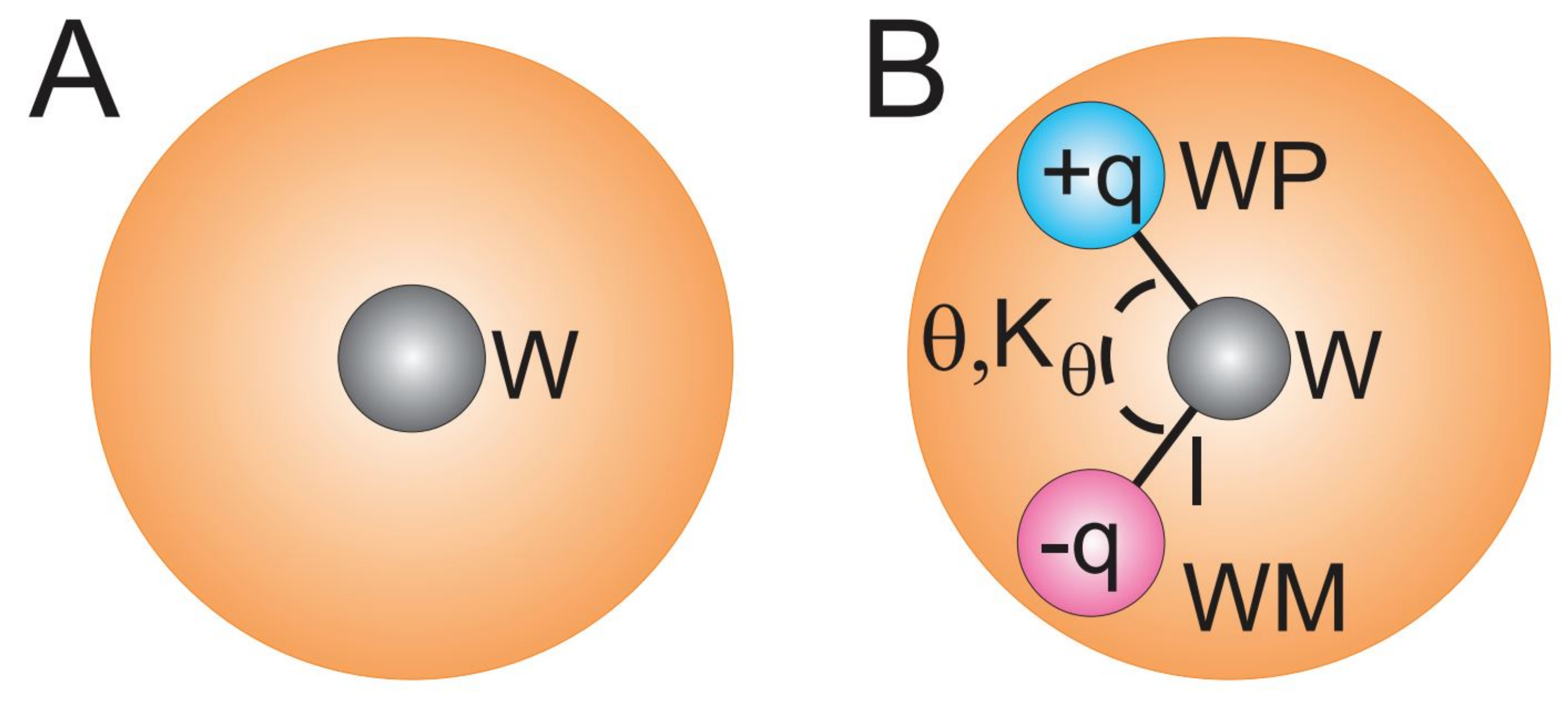}
    \caption{Schematic of the A) original water model and the B) Polarizable water model. The shaded region corresponds to the van der Waal's radii. Reproduced with permission from PLoS Computational Biology.\cite{yesylevskyy2010polarizable}}
  \end{subfigure}
  \caption{A schematic showing the coarse-graining of different forcefields used in this study.}
  \label{fig:schematic}
\end{figure}

We investigate the potential of mean force between poly(lysine) and poly(glutamate) oligomers with 10 residues each.  In most of the simulations all the residues are charged, but we present for comparison one calculation where 5 of the glutamate residues are protonated and 5 of the lysine residues are de-protonated, i.e., the net charge on each is reduced to 6.
We compare results from three different forcefields:   Martini 2.2 with Big Multipole Water\cite{doi:10.1021/jp1019763,doi:10.1021/jp071097f}(BMW-Martini), Martini 2.2 with Polarizable Water\cite{yesylevskyy2010polarizable,doi:10.1021/jp071097f}(POL-Martini) and AMBER ff99sb with TIP3P water\cite{hornak2006comparison}(TIP3P-AMBER). A schematic of the different coarse-grained forcefields is shown in Fig. \ref{fig:schematic}.

Three systems are studied in this work:  with no excess salt but each polyion is neutralized by oppositely charged counter-ions (C$_{NaCl}^{excess}$ = 0 M), 0.27 M excess salt(C$_{NaCl}^{excess}$ = 0.27 M), and no small ions (C$_{NaCl}$=0M).  The solution with C$_{NaCl}^{excess}$ = 0 M would be the true salt-free case, i.e., solutions of salt-free polycations and polyanions are mixed.  The solution with C$_{NaCl}$=0M provides a bench-mark where no small ions are present.  In all cases there is one poly(lysine) and one poly(lysine) molecule with approximate 15,000 atomistic water molecules or 4000 coarse-grained water partilces (one CG water particle corresponds to 4 water molecules).
For C$_{NaCl}^{excess}$ = 0 M and 0.27M there are an additional 10 and 80 particles, respectively, of Cl$^-$ and Na$^+$, for the cases where all the residues are charged.

Simulations are performed using the GROMACS 5.1.4\cite{ABRAHAM201519} package. The Lennard-Jones cutoff is set to 1 nm for the AMBER forcefield and 1.4 nm for the Martini forcefield.  The Particle Mesh Ewald\cite{darden1993particle} method is used to calculate the electrostatic interactions with the following configuration: for the AMBER forcefield, the real cutoff spacing is 1 nm and the fast Fourier transform grid spacing is 0.16 nm; for the Martini forcefield the real cutoff spacing is 1.4 nm and the fast Fourier transform grid spacing is 0.20 nm. 
The Berendsen barostat\cite{doi:10.1063/1.448118} is used to keep the pressure constant, and  the Berendsen Thermostat\cite{doi:10.1063/1.448118} used to keep the temperature constant. 

Initial configurations are created by inserting molecules randomly into in a square-cuboid box of size 12x6x6 nm$^3$ with periodic boundary conditions in all directions.  The energy is minimized using a steepest decent algorithm, and the system is then equilibrated in the NPT ensemble at a pressure of 1 bar. The final configuration obtained from NPT equilibration is used for the pulling simulation in the NVT ensemble.
The two polypeptides are pulled apart along the x-direction to generate multiple windows for the umbrella sampling simulations. For POL-Martini and TIP3P-AMBER forcefields, 40 windows are used for a distance of separation between the central residue of poly(lysine) and poly(glutamate) ($\xi_{LYS-GLU}$) from 0.4 to 3.8 nm. For the BMW-Martini forcefield, 30 windows are used for $\xi_{LYS-GLU}$ between 1 to 3.8 nm. 

For the umbrella sampling production runs, a harmonic force constant of 1000 kJ mol${^{-1}}$ nm$^2$ is applied to constrain the distance of separation between the two polypeptides.   All production runs are done in a NVT ensemble.
Finally, the weighted histogram analysis method\cite{doi:10.1002/jcc.540130812}(WHAM) is employed to obtain the potential of mean force curves from the histograms. The last 75$\%$ of the production runs are used for WHAM analysis. The standard deviation for the PMF curves are computed by using a bootstrapping method in which complete histograms are considered as independent data points. To ensure that the system is equilibrated, the PMF obtained from the first 25$\%$ of the production run is compared to that obtained from the last 75$\%$ of the simulation run. The two potential of mean curves were within less than half a standard deviation of each other.

Using the method thus described, PMF curves are obtained for two temperatures, at a lower temperature, T$_1$ and at a higher temperature, T$_2$. For the BMW-Martini forcefield, T$_1$ = 290 K and T$_2$ = 310 K. For POL-Martini and TIP3P-AMBER, T$_1$ = 280 K and T$_2$ = 320 K. Assuming that the energy and entropy of association is constant between the temperature of T$_1$ and T$_2$, the PMF curves obtained from umbrella sampling are decomposed into energetic($\Delta U(\xi$)) and entropic($\Delta S(\xi$)) contributions at a given distance of separation using the equations:
    \begin{equation}
        \Delta S(\xi) = - \frac{\Delta A(\xi, T_2) - \Delta A(\xi, T_1)}{(T_2 - T_1)}
       \label{eq:1}
    \end{equation}
\begin{equation}
    \Delta U(\xi) = \Delta A(\xi) + T\Delta S(\xi)
    \label{eq:2}
\end{equation}
Here $\Delta$A is the Hemholtz free energy which is numerically equal to the value of the shifted PMF curve. The standard deviation for $\Delta A$ is calculated by using bootstrap analysis. The error bars shown in the plots correspond to one standard deviation of the quantity of interest.

\section{Results and Discussion}

\subsection{Potential of mean-force}

\begin{figure}[h!]

  \centering
  \begin{subfigure}[b]{0.45\linewidth}
    \includegraphics[trim={3.2cm 9cm 1.2cm 9cm},clip,width=\linewidth]{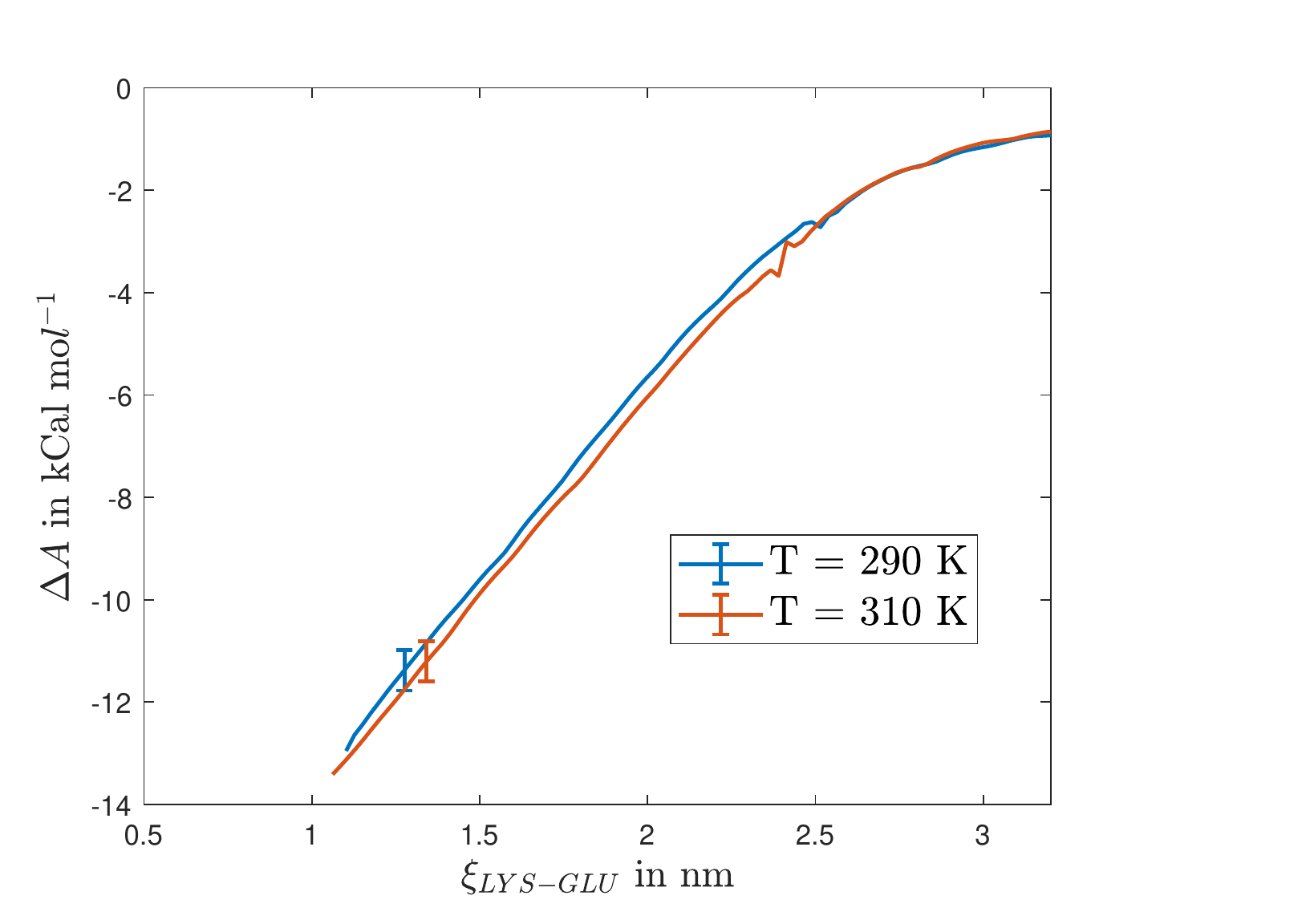}
    \caption{BMW-Martini, C$_{NaCl}^{excess}$ = 0 M}
  \end{subfigure}
  \begin{subfigure}[b]{0.45\linewidth}
    \includegraphics[trim={3.2cm 9cm 1.2cm 9cm},clip,width=\linewidth]{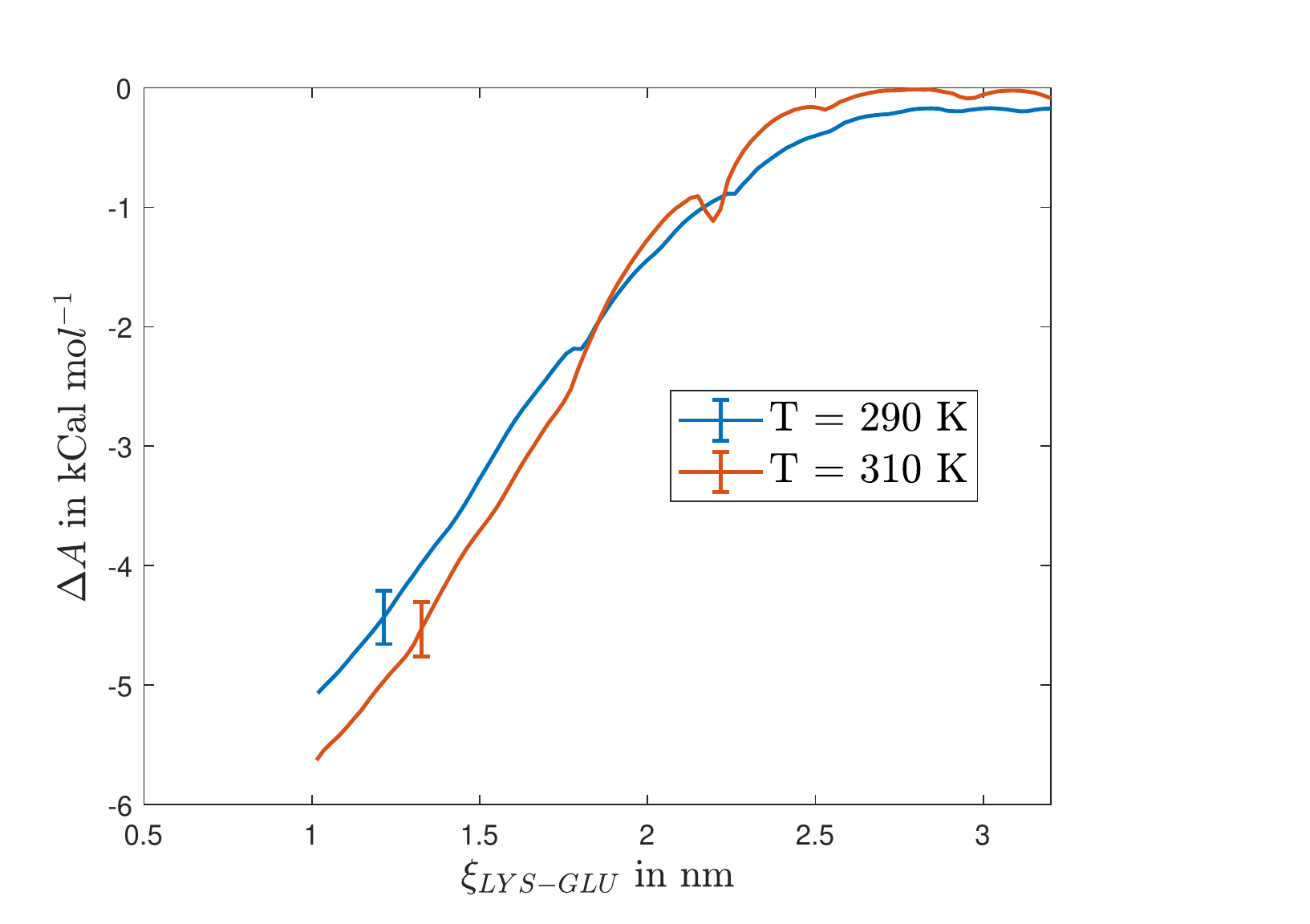}
    \caption{BMW-Martini, C$_{NaCl}^{excess}$ = 0.27 M}
  \end{subfigure}
  \begin{subfigure}[b]{0.45\linewidth}
    \includegraphics[trim={3.2cm 9cm 1.2cm 9cm},clip,width=\linewidth]{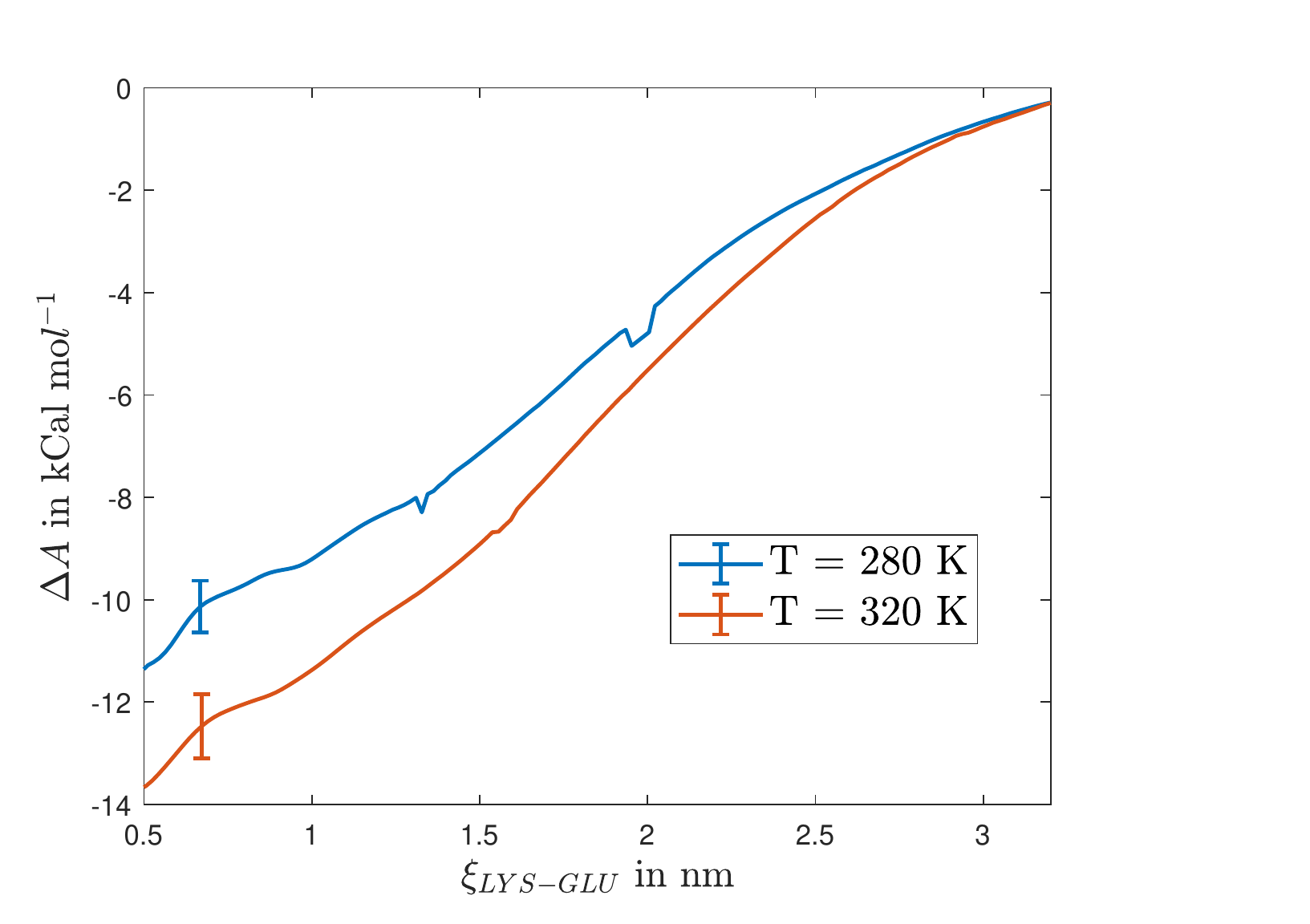}
    \caption{POL-Martini, C$_{NaCl}^{excess}$ = 0 M}
  \end{subfigure}
  \begin{subfigure}[b]{0.45\linewidth}
    \includegraphics[trim={3.2cm 9cm 1.2cm 9cm},clip,width=\linewidth]{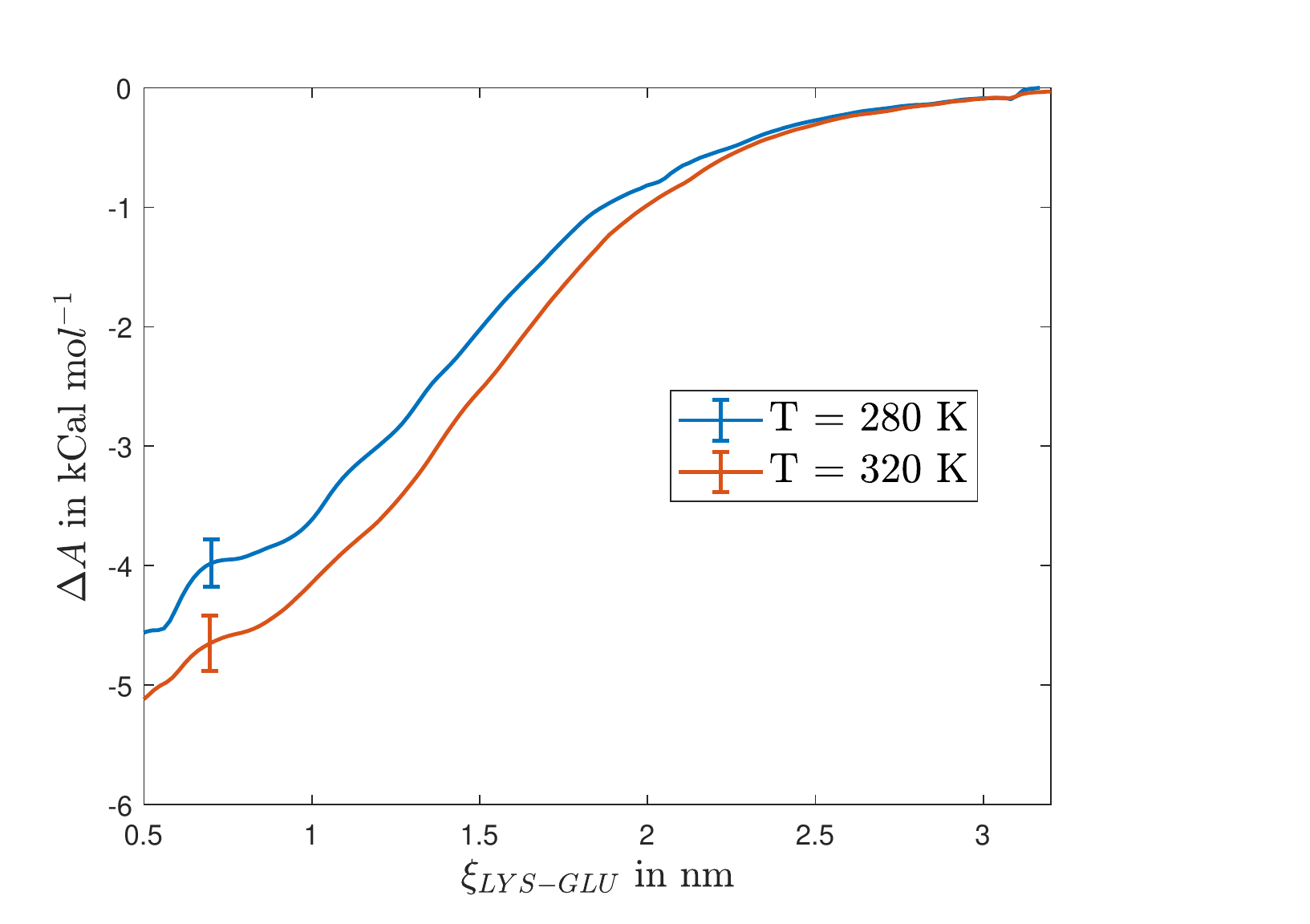}
    \caption{POL-Martini, C$_{NaCl}^{excess}$ = 0.27 M}
  \end{subfigure}
  \begin{subfigure}[b]{0.45\linewidth}
    \includegraphics[trim={3.2cm 9cm 1.2cm 9cm},clip,width=\linewidth]{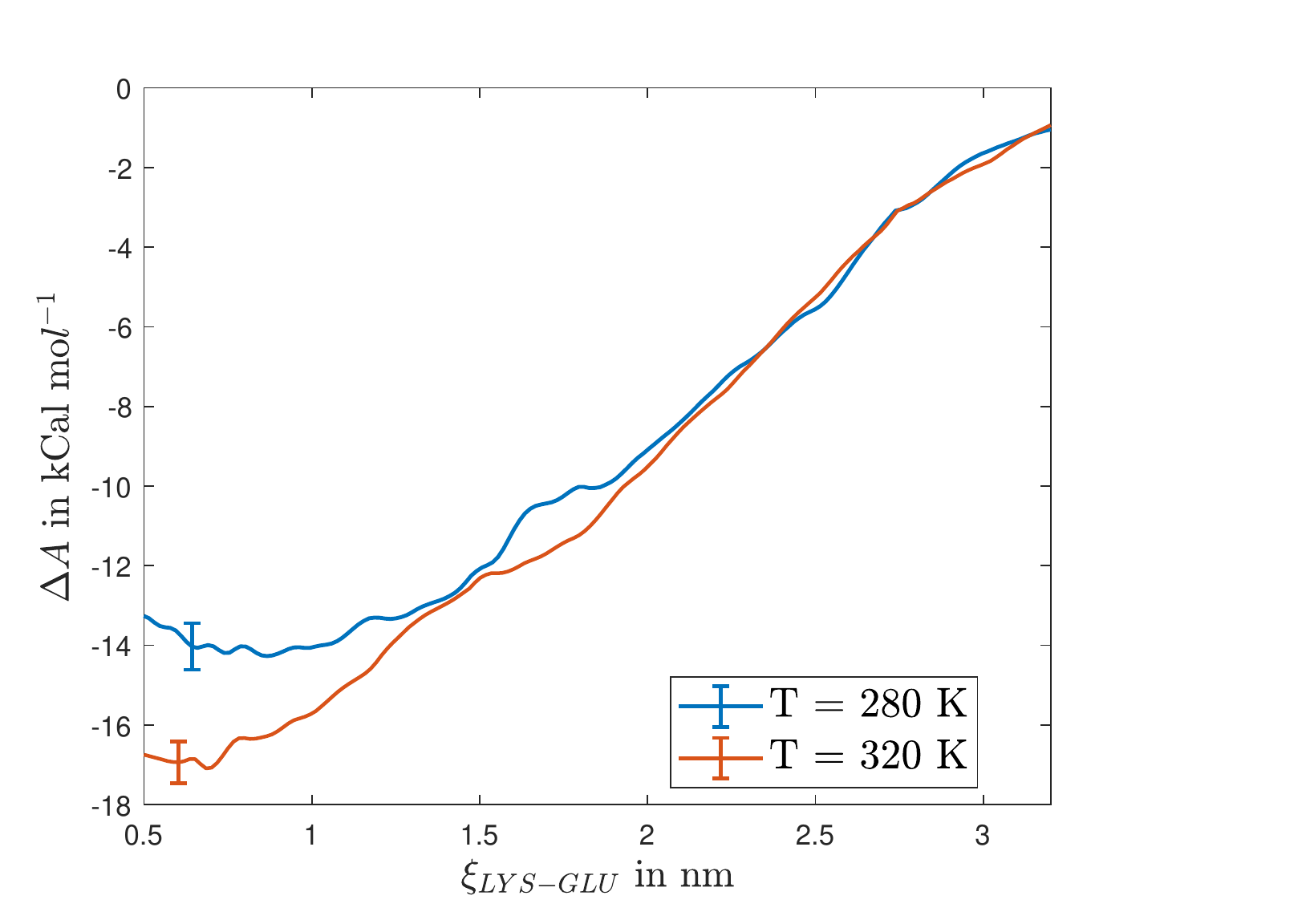}
    \caption{TIP3P-AMBER, C$_{NaCl}^{excess}$ = 0 M}
  \end{subfigure}
  \begin{subfigure}[b]{0.45\linewidth}
    \includegraphics[trim={3.2cm 9cm 1.2cm 9cm},clip,width=\linewidth]{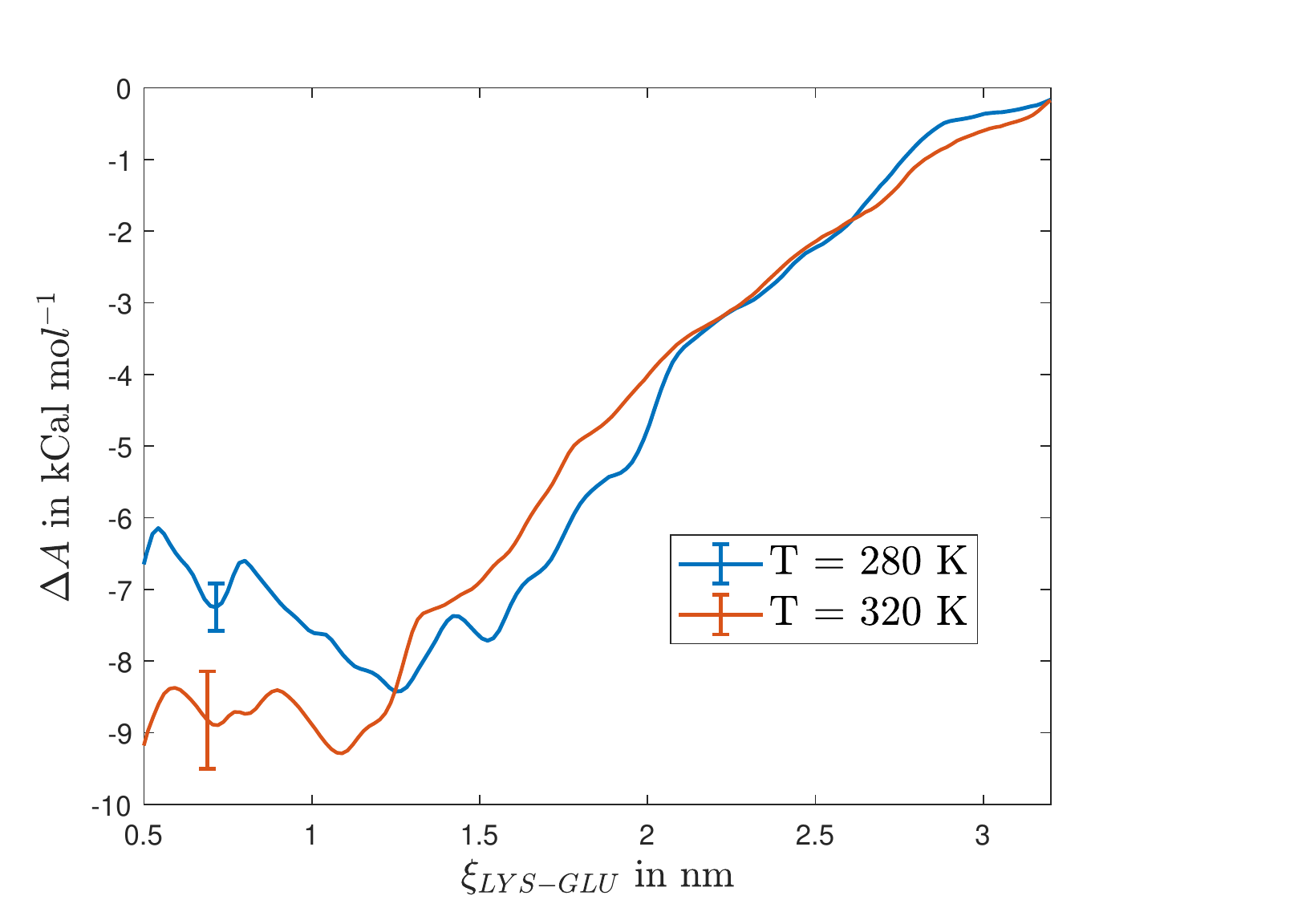}
    \caption{TIP3P-AMBER, C$_{NaCl}^{excess}$ = 0.27 M}
  \end{subfigure}
  \caption{Potential of mean force between poly(lysine) and poly(glutamate).}
  \label{fig:pmf}
\end{figure}

The potential of mean curves with the three forcefields is shown in figure~\ref{fig:pmf} for C$_{NaCl}^{excess}$ = 0 M and 0.27M, at two different temperatures in each case.  As expected, there is a favorable (negative) free energy of association in all cases.  The qualitative behavior is the same in all force fields although there are quantitative differences.  In particular the magnitude of association is stronger in the atomistic model.  
 
To compare the strength of association we tabulate, in table~\ref{table:pmf}, the free energy of association, defined as the difference in free energy between the value at 1nm and the value far away.   For the system with no excess salt and the system with excess salt at T$_1$, both POL-Martini and BMW-Martini underestimate the free energy of association by $\approx$ 3 kcal mol$^{-1}$ as compared to AMBER. For the system with excess salt at T$_2$, the free energy of association for the AMBER forcefield is 5 kcal mol$^{-1}$ more favorable than both POL-Martini and BMW-Martini. This quantitative difference in the free energy can be possibly attributed to the mapping protocol of representing multiple beads of atomistic forcefield to a considerably smaller number of molecular sites.\cite{doi:10.1021/ct501063a}. Another quantitative discrepancy that we observe is the change in free energy with the addition of salt - addition of 0.27 M of excess salt decreases the free energy of association by $\approx$ 7 kcal mol$^{-1}$ for the coarse grained forcefields and $\approx$ 3 kcal mol$^{-1}$ for the atomistic forcefield. 
\begin{table}[]
\begin{tabular}{lllll}
\hline

System                               & \multicolumn{4}{c}{\textbf{$\Delta$A in kcal mol$^{-1}$}}                                \\ 
                                     & \multicolumn{2}{l}{C$_{NaCl}^{excess}$ = 0 M} & \multicolumn{2}{l}{C$_{NaCl}^{excess}$ = 0.27 M} \\
                                     & T$_1$             & T$_2$            & T$_1$              & T$_2$              \\\hline
Martini 2.2 with Big Multipole Water & -12.64            & -13.55            & -5.00              & -5.63             \\ 
Martini 2.2 with Polarizable Water   & -9.22            & -11.37           & -3.71               & -4.161              \\
AMBER ff99sb with TIP3P Water        & -14.27            & -16.96           & -7.07              & -10.19             
\end{tabular}
\caption{Value of the free energy for $\xi_{LYS-GLU}$=1nm for different forcefields at 1nm T$_1$ = 290 K for BMW-Martini and 280 K for POL-Martini and TIP3P-AMBER. T$_2$ = 310 K for BMW-Martini and 320 K for POL-Martini and TIP3P-AMBER}
\label{table:pmf}
\end{table}

\subsection{Decomposition into entropic and energetic contributions}

\begin{figure}[h!]
  \centering
  \begin{subfigure}[b]{0.45\linewidth}
    \includegraphics[trim={3.2cm 9cm 1.2cm 9cm},clip,width=\linewidth]{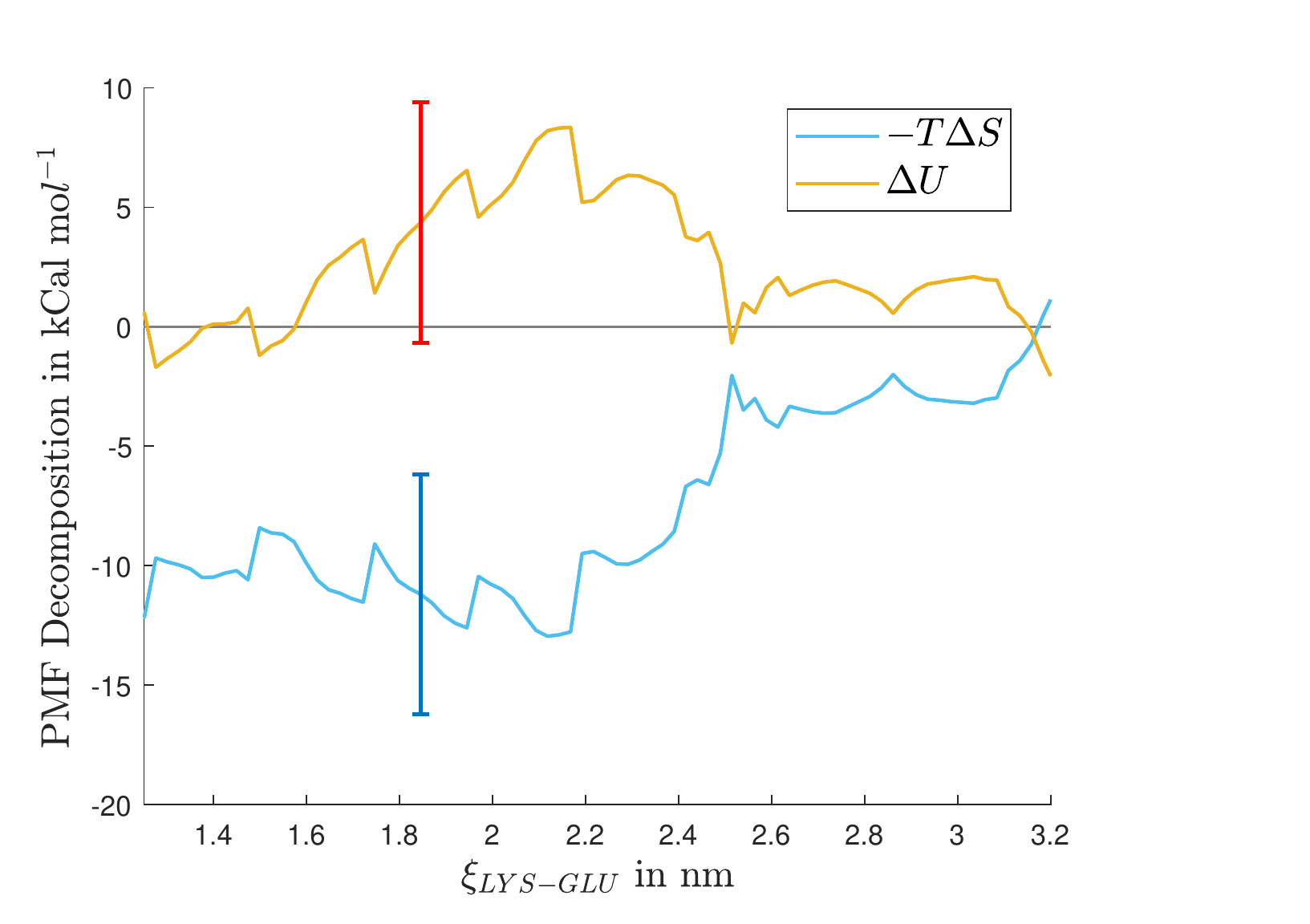}
    \caption{BMW-Martini, C$_{NaCl}^{excess}$ = 0 M}
  \end{subfigure}
  \begin{subfigure}[b]{0.45\linewidth}
    \includegraphics[trim={3.2cm 9cm 1.2cm 9cm},clip,width=\linewidth]{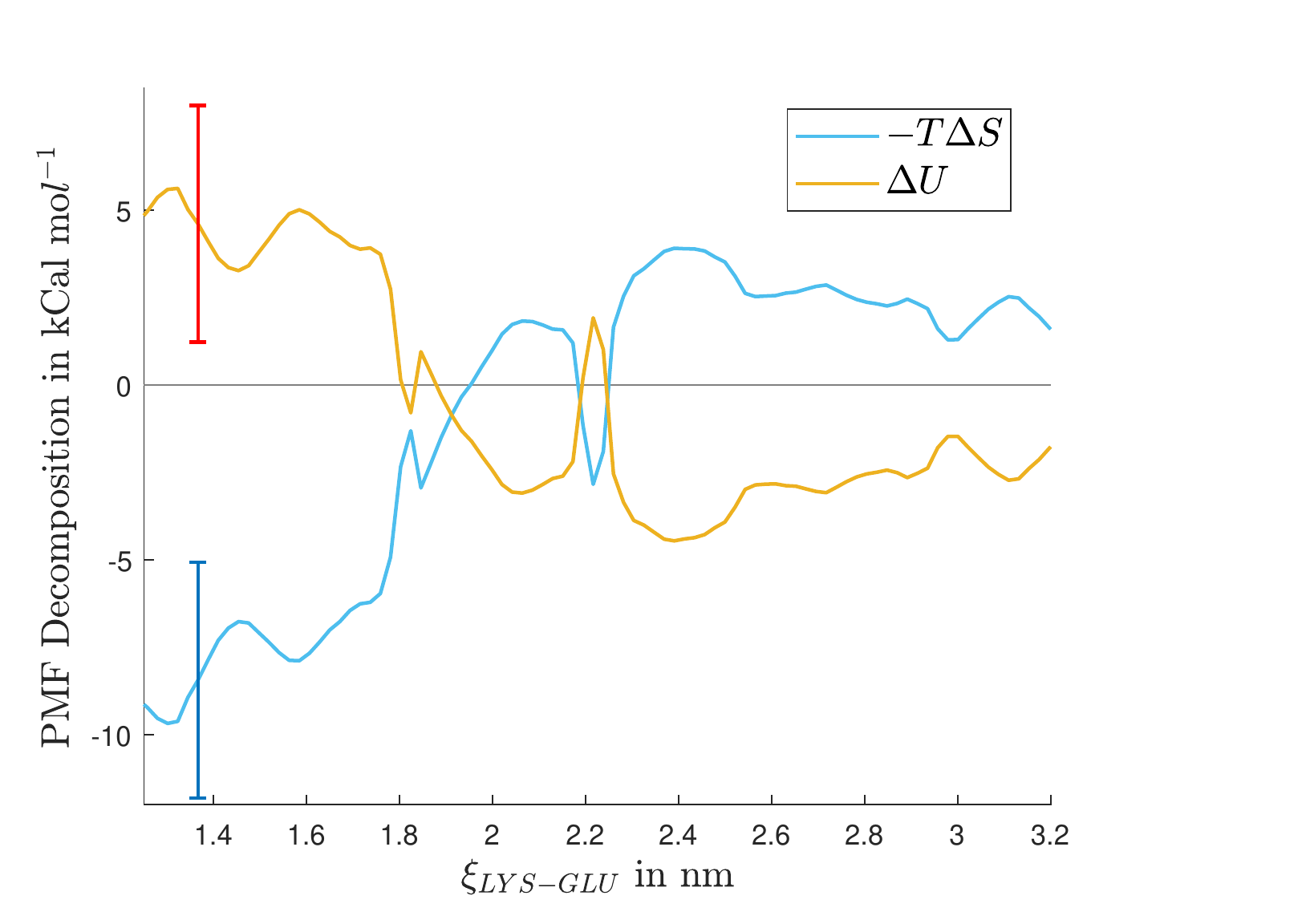}
    \caption{BMW-Martini, C$_{NaCl}^{excess}$ = 0.27 M}
  \end{subfigure}
  \begin{subfigure}[b]{0.45\linewidth}
    \includegraphics[trim={3.2cm 9cm 1.2cm 9cm},clip,width=\linewidth]{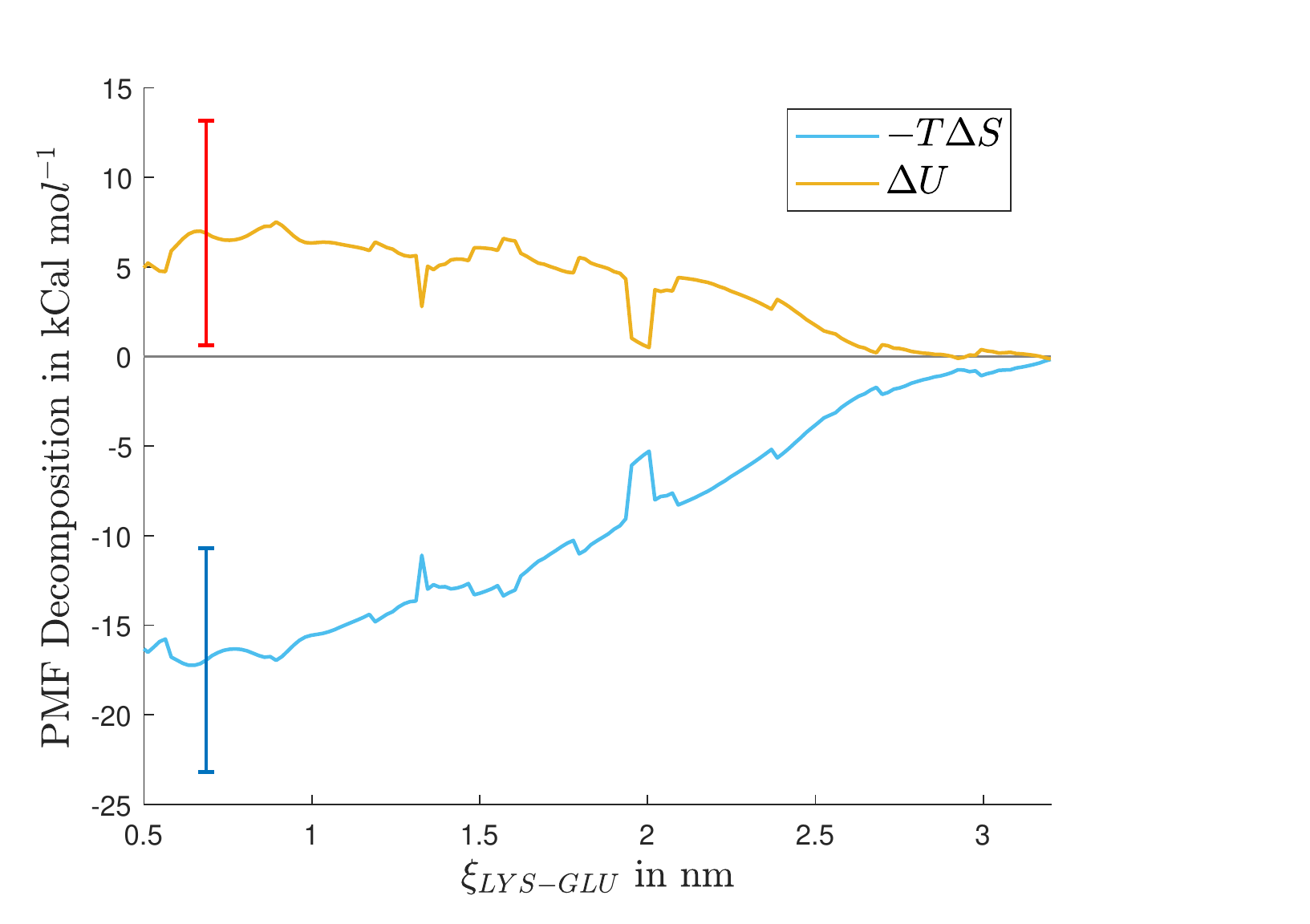}
    \caption{POL-Martini, C$_{NaCl}^{excess}$ = 0 M}
  \end{subfigure}
  \begin{subfigure}[b]{0.45\linewidth}
    \includegraphics[trim={3.2cm 9cm 1.2cm 9cm},clip,width=\linewidth]{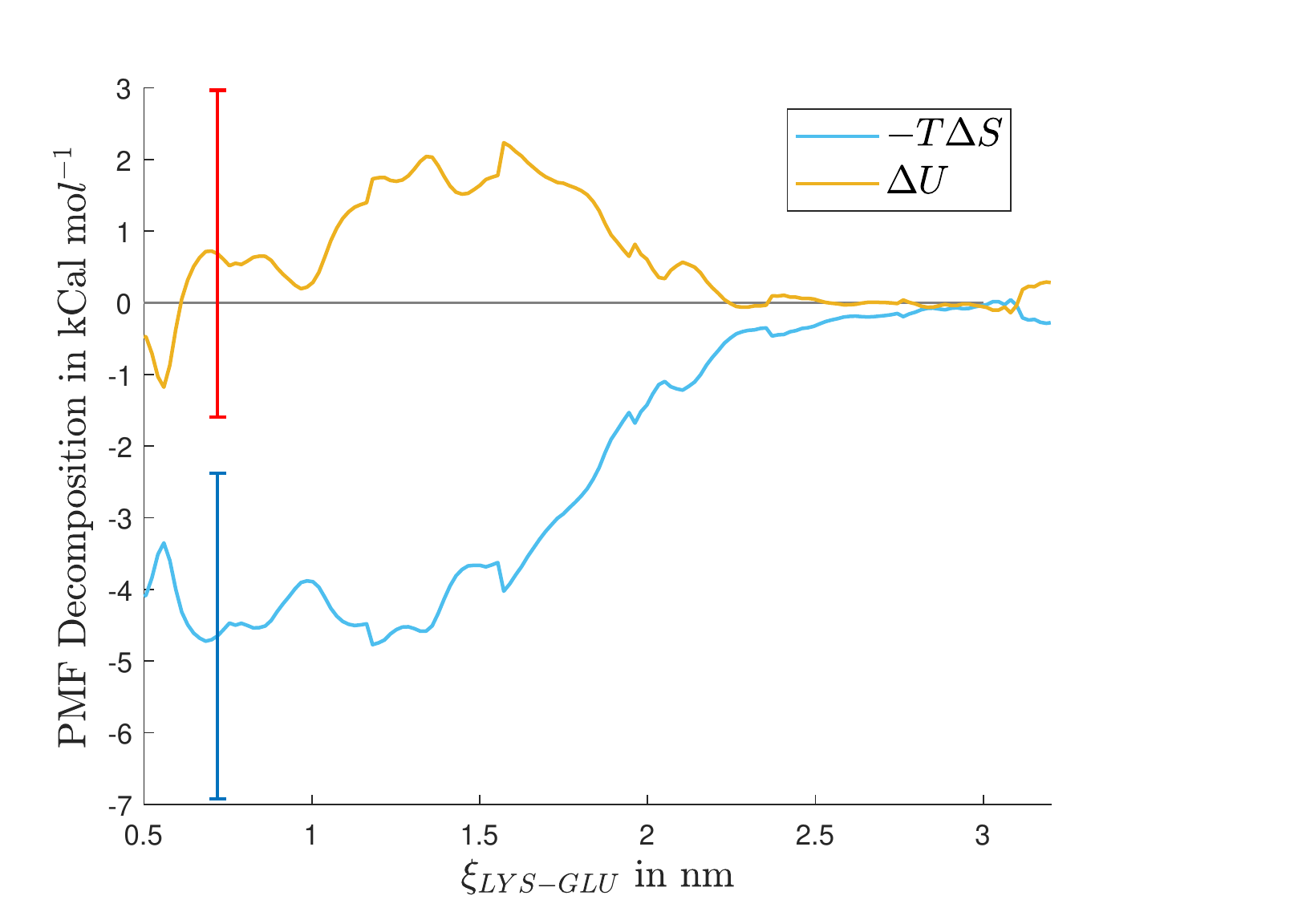}
    \caption{POL-Martini, C$_{NaCl}^{excess}$ = 0.27 M}
  \end{subfigure}
  \begin{subfigure}[b]{0.45\linewidth}
    \includegraphics[trim={3.2cm 9cm 1.2cm 9cm},clip,width=\linewidth]{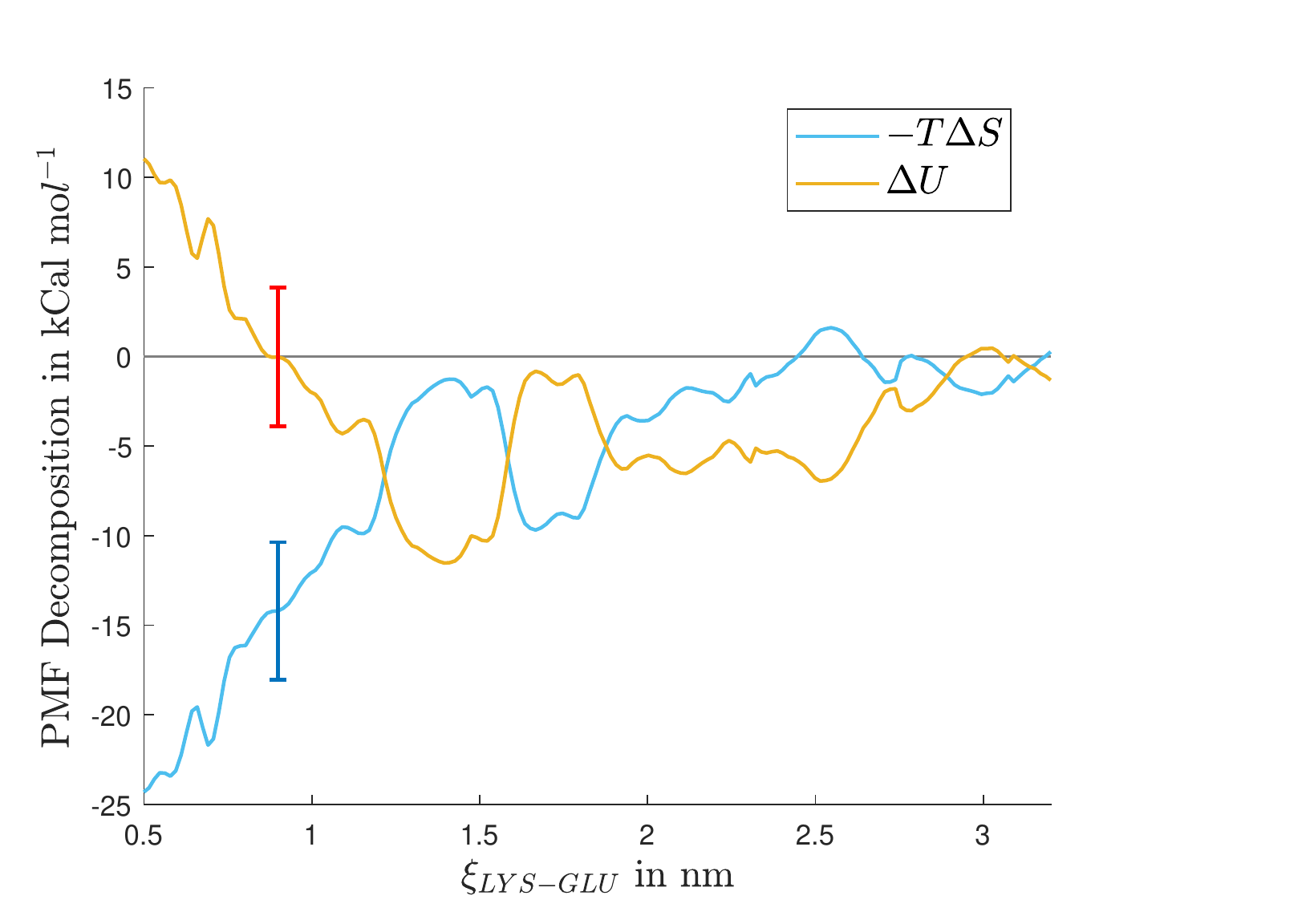}
    \caption{TIP3P-AMBER, C$_{NaCl}^{excess}$ = 0 M}\label{fig:3e}
  \end{subfigure}
  \begin{subfigure}[b]{0.45\linewidth}
    \includegraphics[trim={3.2cm 9cm 1.2cm 9cm},clip,width=\linewidth]{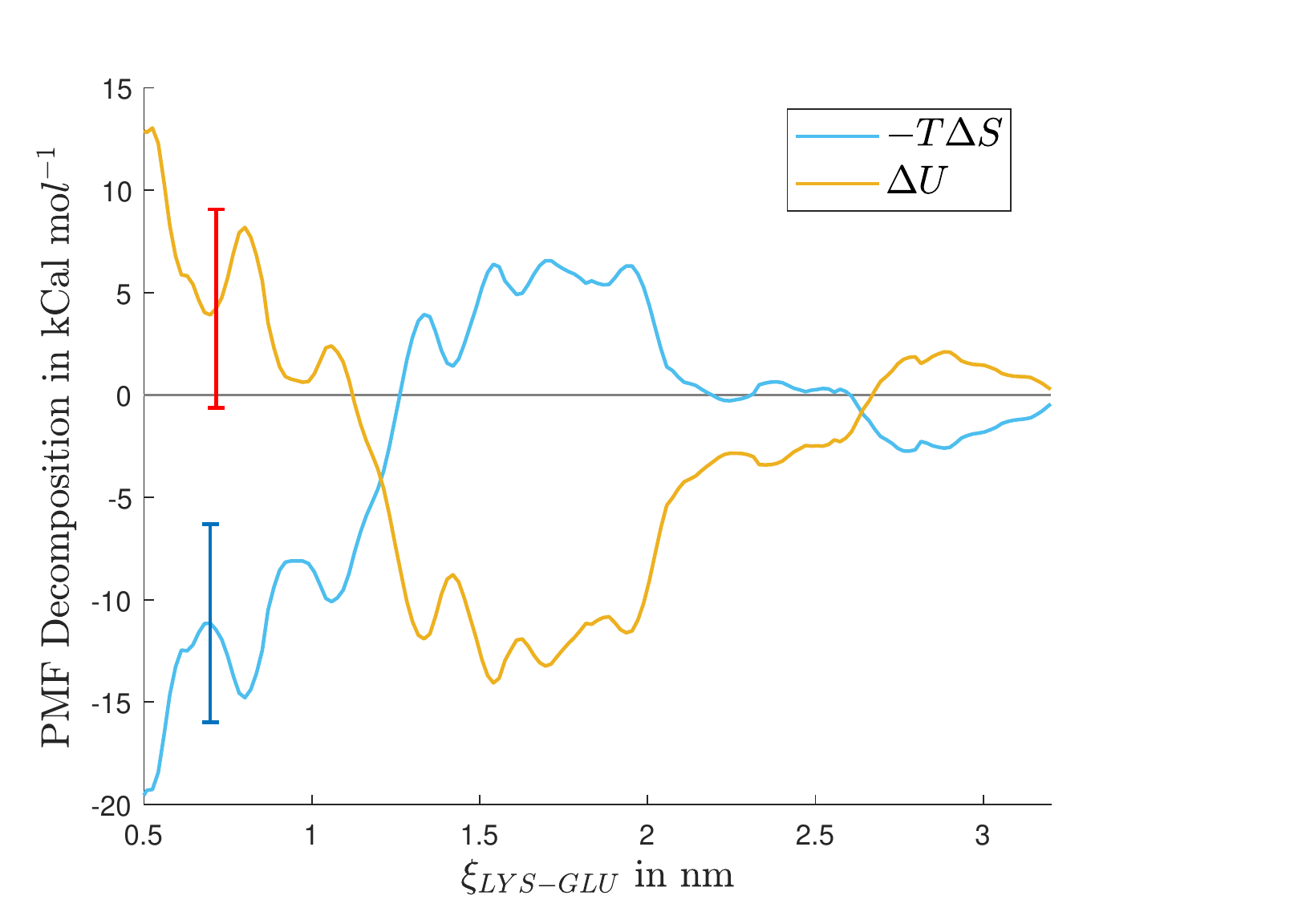}
    \caption{TIP3P-AMBER, C$_{NaCl}^{excess}$ = 0.27 M}
  \end{subfigure}
  \caption{Entropic and energetic contributions to the potential of mean force.}
  \label{fig:decompositions}
\end{figure}

The entropic and energetic contribution to the PMF are shown in figure~\ref{fig:decompositions}.   Results from the three forcefields are consistent in that they reveal that the association is strongly entropically favored at small separations. For TIP3P-AMBER, the driving force switches from being entropically-driven to energetically-driven at separations of $\approx$ 1.5-2nm.  

For the TIP3P-AMBER case, the PMF has a stronger energetic contribution at larger separations, especially at the higher salt concentration.  This is true in the BMW-Martini result at high salt but is absent in the POL-Martini results.  The entropy, energy, and free energy all decrease as the salt concentration is increased.  This is consistent with past experimental and theoretical works on this phenomenon\cite{van2011polyelectrolyte,doi:10.1021/jp062264z,B605695D,doi:10.1021/jacs.5b11878}. 

The major distinction we find however is the endothermic nature of complexation in the absence of excess salt, and the complexation is entropically driven.  A recent experimental study by Fu and Schlenoff \cite{doi:10.1021/jacs.5b11878} suggests that the enthalpic contribution for complex coacervation arises from the changes in water perturbation.  This is unlikely in our model, however, because the qualitative behavior (the driving factors of complexation) is insensitive to the water model.

\begin{figure}[h!]
  \centering

  \begin{subfigure}[b]{0.45\linewidth}
    \includegraphics[trim={3.2cm 9cm 1.2cm 9cm},clip,width=\linewidth]{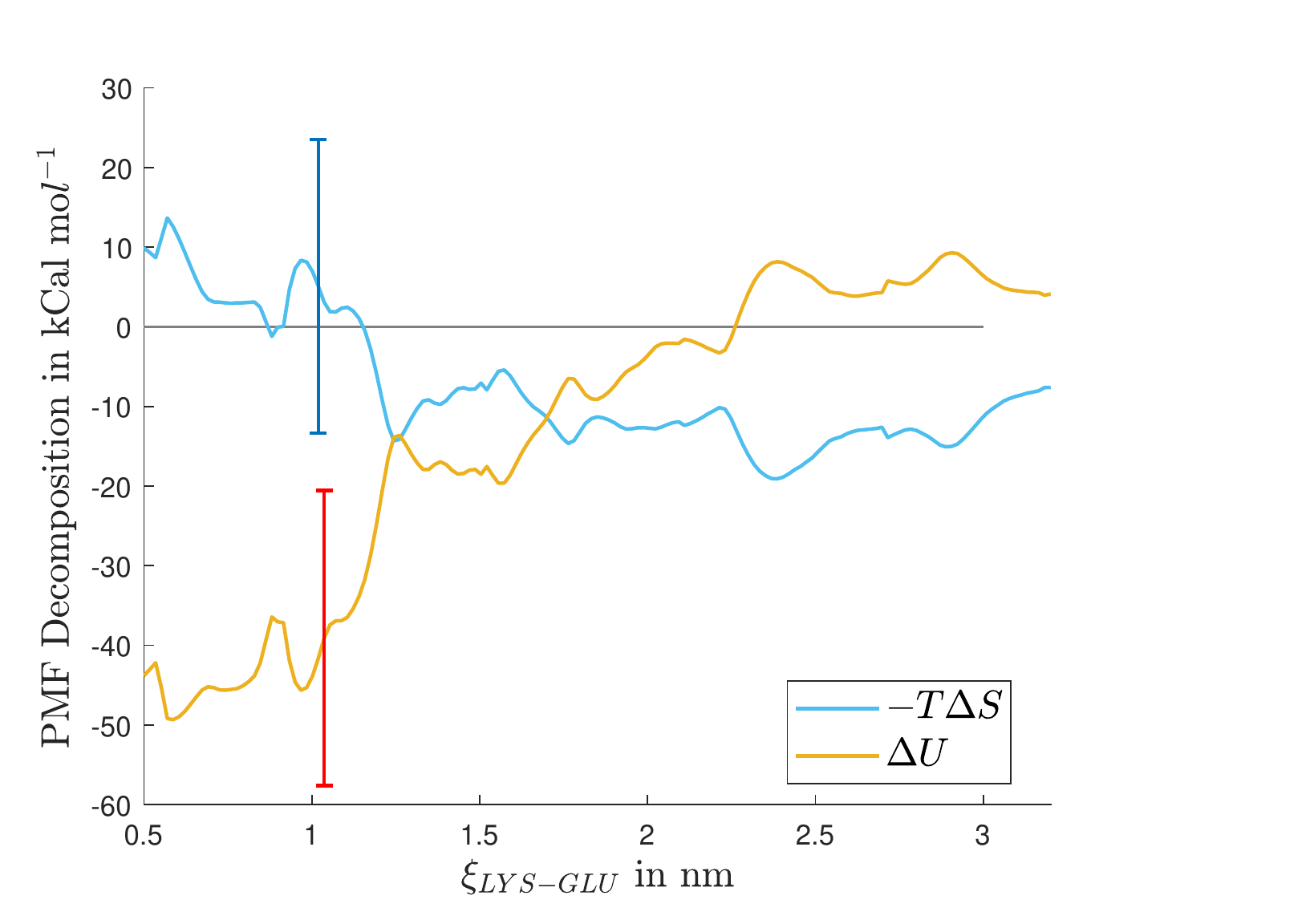}
  \end{subfigure}

  \caption{Entropic and energetic contributions to the potential of mean force with TIP3P-AMBER and $C_{NaCl}$=0M. There are no counterions or excess salt present in this system.}
    \label{fig:5}
\end{figure}

The removal of all small ions makes the driving force energetic.  Figure~\ref{fig:5} depicts the entropic and energetic contributions for C$_{NaCl}$=0M.  At all temperatures the association is significantly stronger in the absences of small ions; the magnitude of the free energy of complexation is $\approx$ 42.5 kcal mol$^{-1}$.  This suggests that the polyion electrostatic interactions are screened by the small ions even at very short distances.  The main result, however, is that the driving force is energetic, which means that just the presence of neutralizing small ions is sufficient to change the nature of the thermodynamic driving force.  This strongly indicates that the favorable entropy for complexation predominantly comes from the counterions and the excess salt in the systems. 

\begin{figure}[h!]
  \centering

  \begin{subfigure}[b]{0.45\linewidth}
    \includegraphics[trim={3.2cm 9cm 1.2cm 9cm},clip,width=\linewidth]{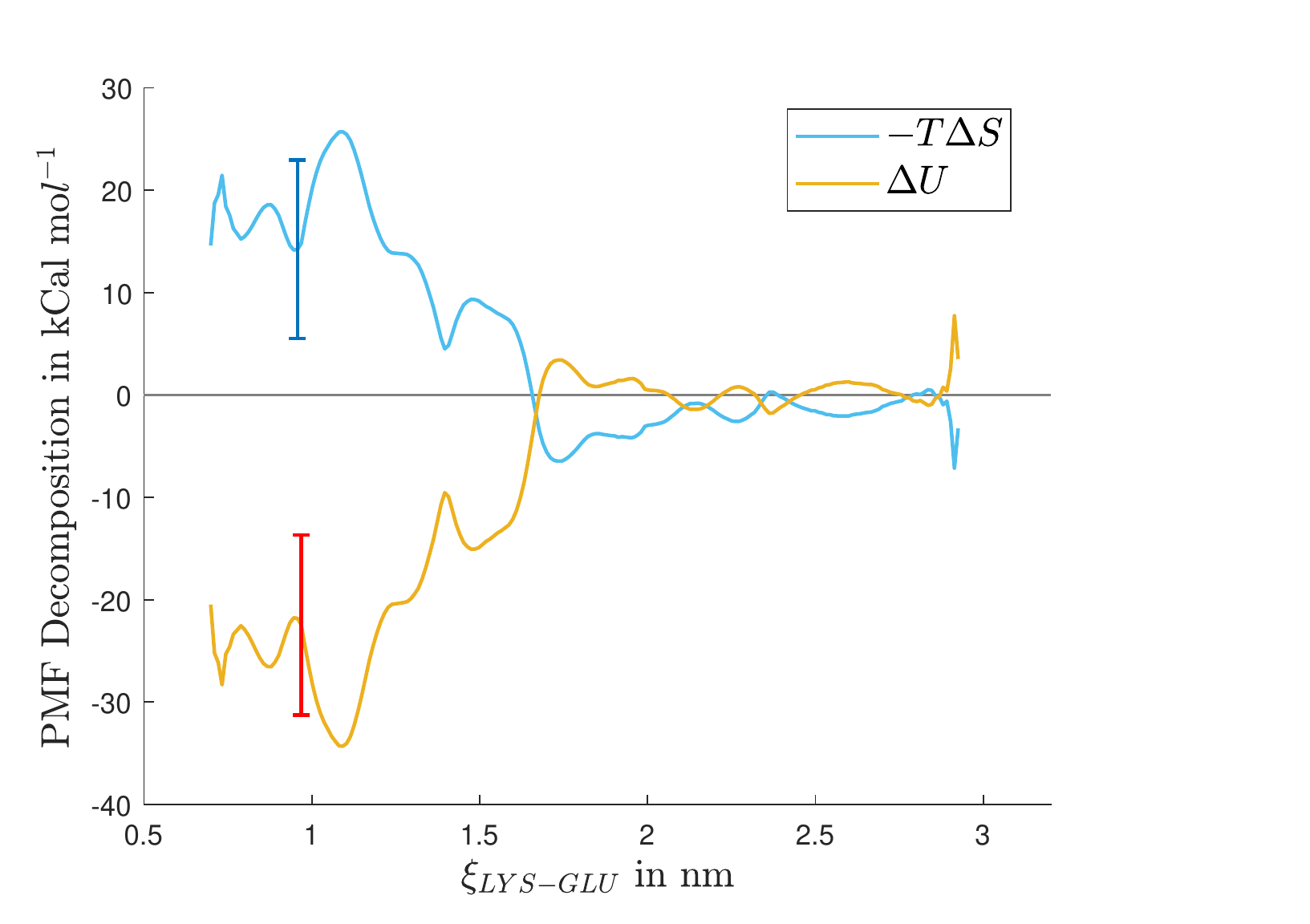}
  \end{subfigure}

  \caption{Entropic and energetic contributions to the potential of mean force with TIP3P-AMBER and $C^{excess}_{NaCl}$=0M, for 11-mer peptides with 6 charged groups.}
    \label{fig:11mer}
\end{figure}

The driving force is also energetic if the charge on the peptides is reduced.  Figure~\ref{fig:11mer} depicts the entropic and energetic contributions for a 11-mer peptide with 6 charged groups in salt-free conditions, with the TIP3P-AMBER force field.  With the reduced charge density of the polyions, and the correspondingly smaller number of counterions, the driving force is energetic, compared to the entropic PMF observed for fully charged peptides in salt-free solution.  This further implies that the entropic driving force arises from contributions from the counterions.

\subsection{Contribution of counterions in the entropy of complexation}

\begin{figure}[h!]
  \centering
  \begin{subfigure}[b]{0.8\linewidth}
  \centering
    \includegraphics[trim={0cm 2cm 1cm 2cm},clip,width=.5\linewidth]{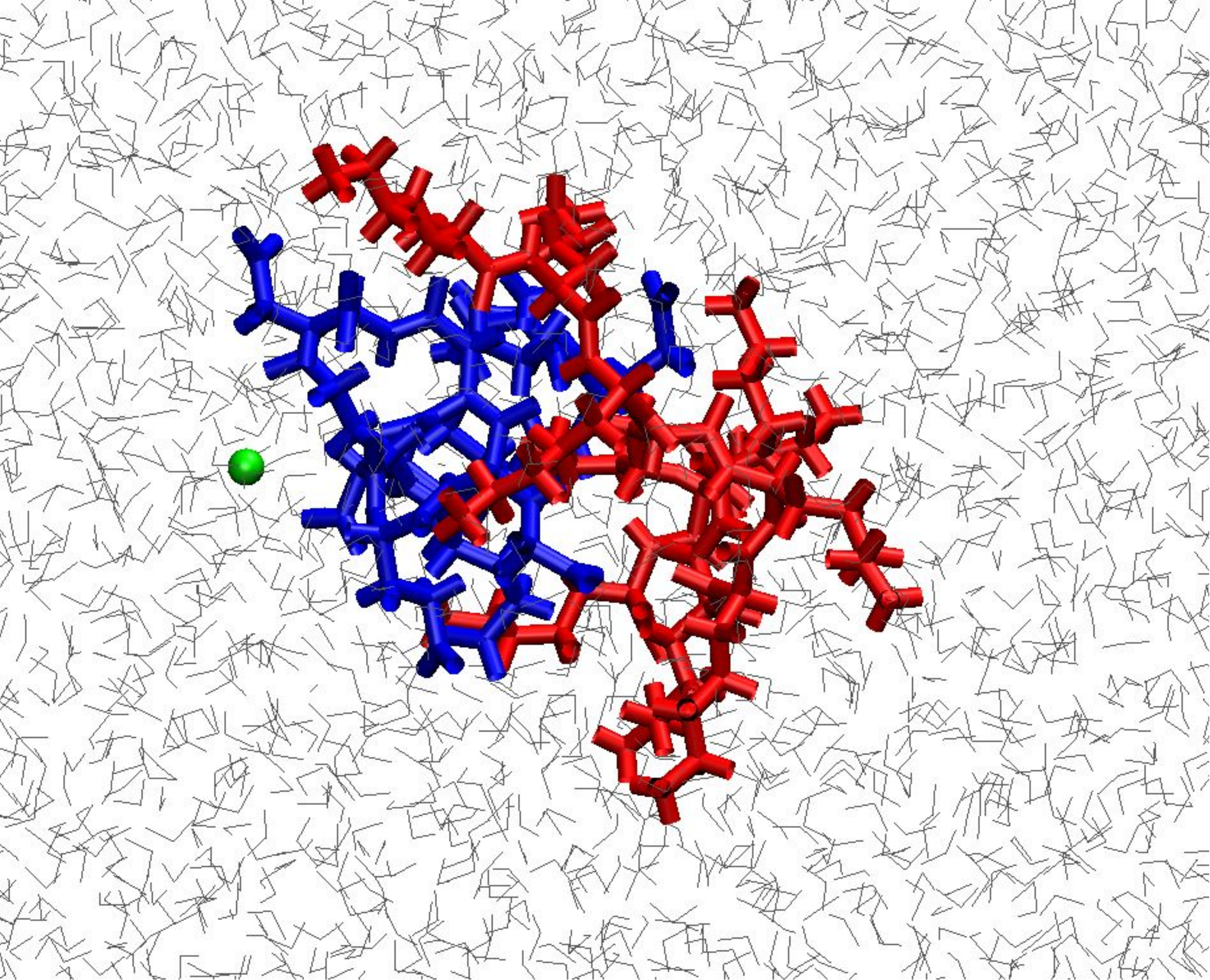}
    \caption{Poly(lysine) and Poly(glutamate), $\xi_{LYS-GLU}$ = 1 nm}\label{fig:6a}
  \end{subfigure}
  \begin{subfigure}[b]{0.4\linewidth}
    \includegraphics[width=\linewidth]{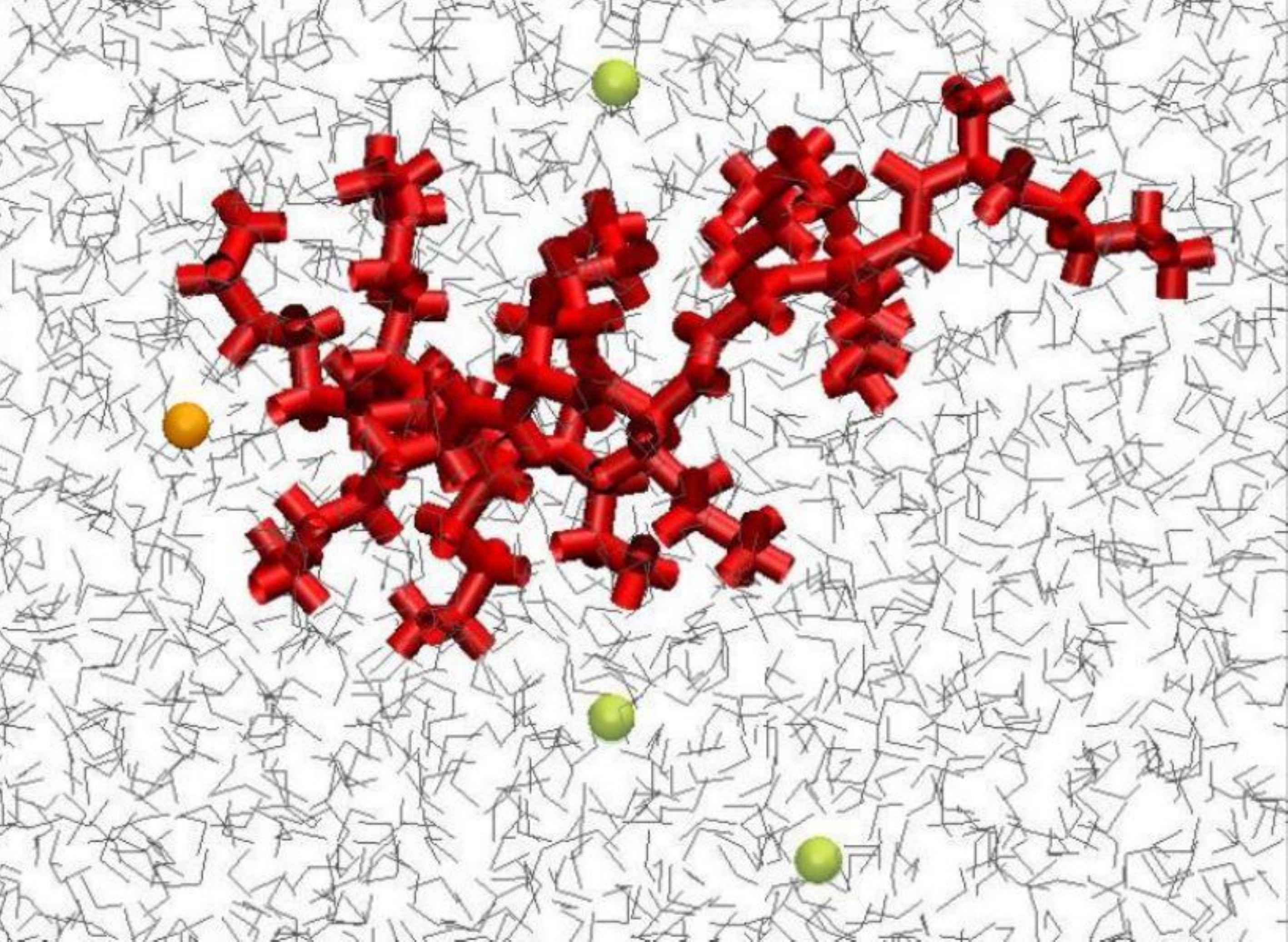}
    \caption{Poly(lysine), $\xi_{LYS-GLU}$ = 3.5 nm}\label{fig:6b}
  \end{subfigure}
  \begin{subfigure}[b]{0.4\linewidth}
    \includegraphics[trim={0cm 1.4cm 0cm .6cm},clip,width=\linewidth]{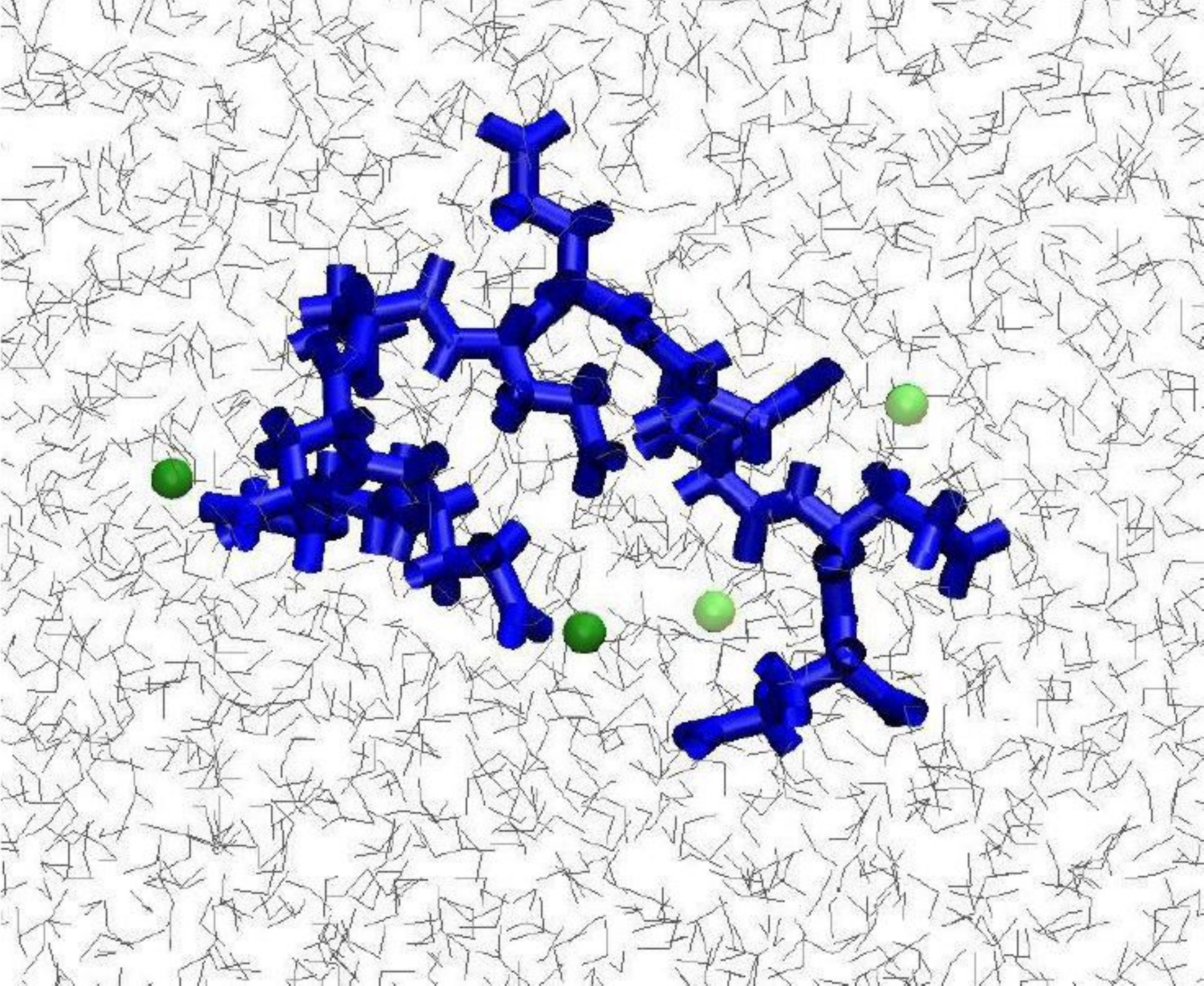}
    \caption{Poly(glutamate), $\xi_{LYS-GLU}$ = 3.5 nm}\label{fig:6c}
  \end{subfigure}
  \caption{Snapshots of the poly(lysine) (red) and poly(glutamate) (blue) when they are complexed (a) and when they are separated (b \& c). Na$^+$ is colored green and Cl$^-$ is colored yellow. Ions within the second shell of the charged groups of the polypeptides are darkly shaded.}
  \label{fig:snapshots}
\end{figure}

The role of counterions in the entropy of complexation is subtle.  Figures~\ref{fig:6a}, \subref{fig:6b} and \subref{fig:6c} depict snapshots of the polyions in the case where $\xi_{LYS-GLU}$ = 1.0 nm (part a) and $\xi_{LYS-GLU}$ = 3.5 nm (parts b and c).  The separation between polyions is constrained using an umbrella potential as in the PMF calculation.  
It is evident from the snapshots that there is a substantial increase in the coordination number of the counterions for both polypeptides when they are not interacting with each other.  The counterions are not statically constrained, of course, and their diffusion coefficient is not significantly reduced.  Note also that the fraction of all counterions within the second shell of the charged groups of the polyion is only 0.3-0.4.

\begin{figure}[h!]
  \centering 
  \begin{subfigure}[b]{0.45\linewidth}
    \includegraphics[trim={3.2cm 9cm 1.2cm 9cm},clip,width=\linewidth]{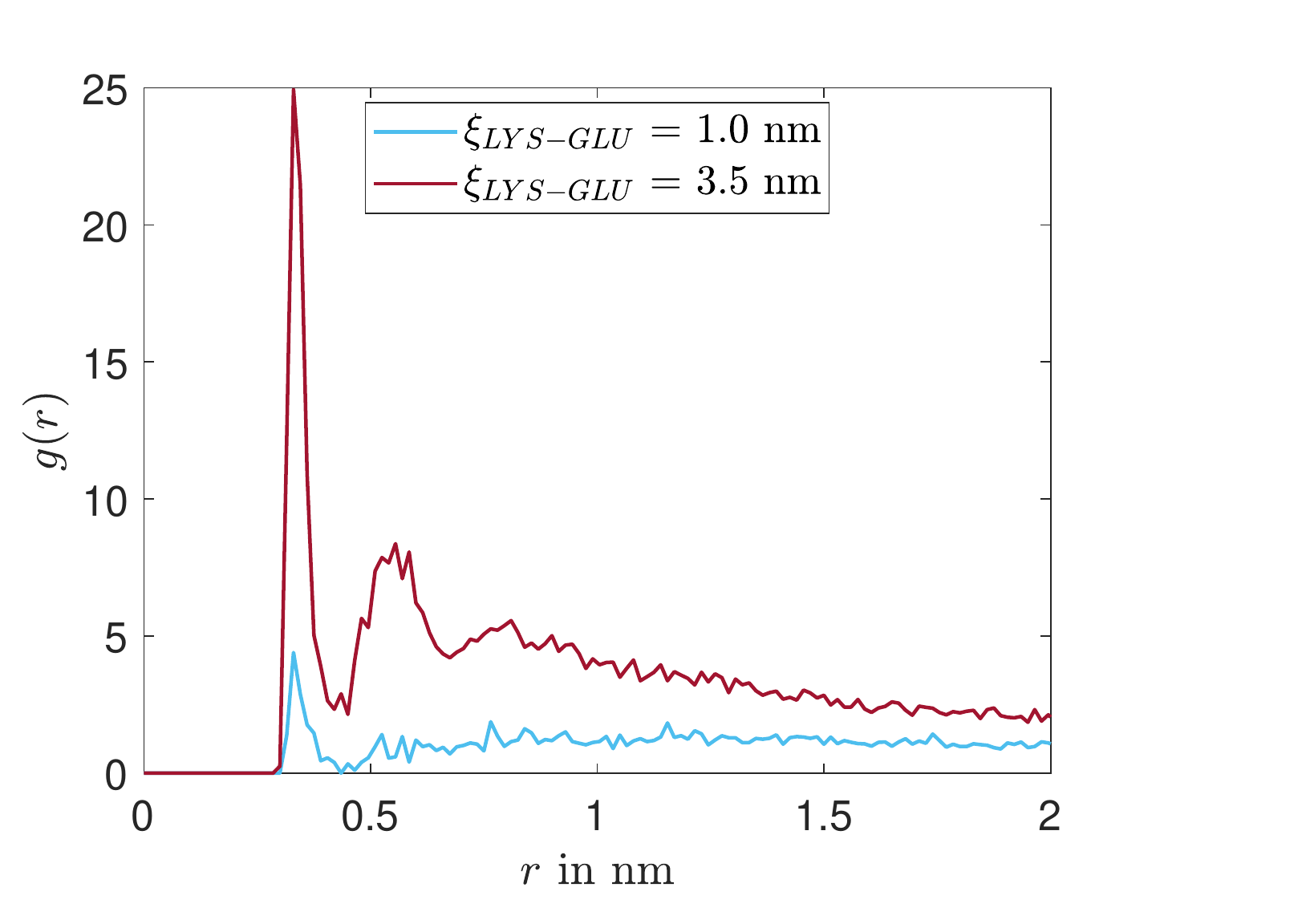}
    \caption{Poly(lysine)-Cl$^-$}\label{fig:7a}
  \end{subfigure}
  \begin{subfigure}[b]{0.45\linewidth}
    \includegraphics[trim={3.2cm 9cm 1.2cm 9cm},clip,width=\linewidth]{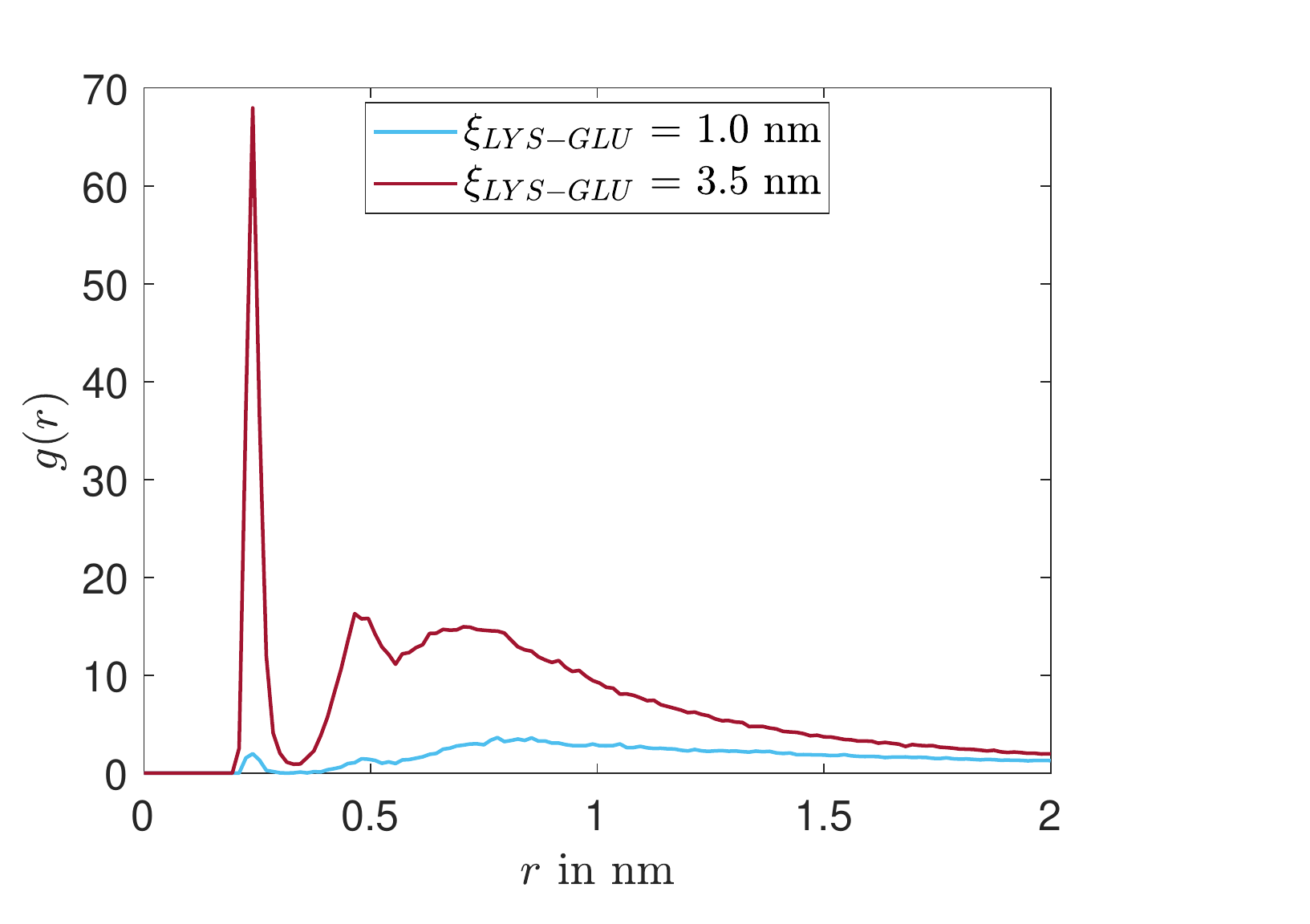}
    \caption{Poly(glutamate)-Na$^+$}\label{fig:7b}
  \end{subfigure}
  \begin{subfigure}[b]{0.5\linewidth}
    \includegraphics[trim={3.2cm 9cm 1.2cm 9cm},clip,width=\linewidth]{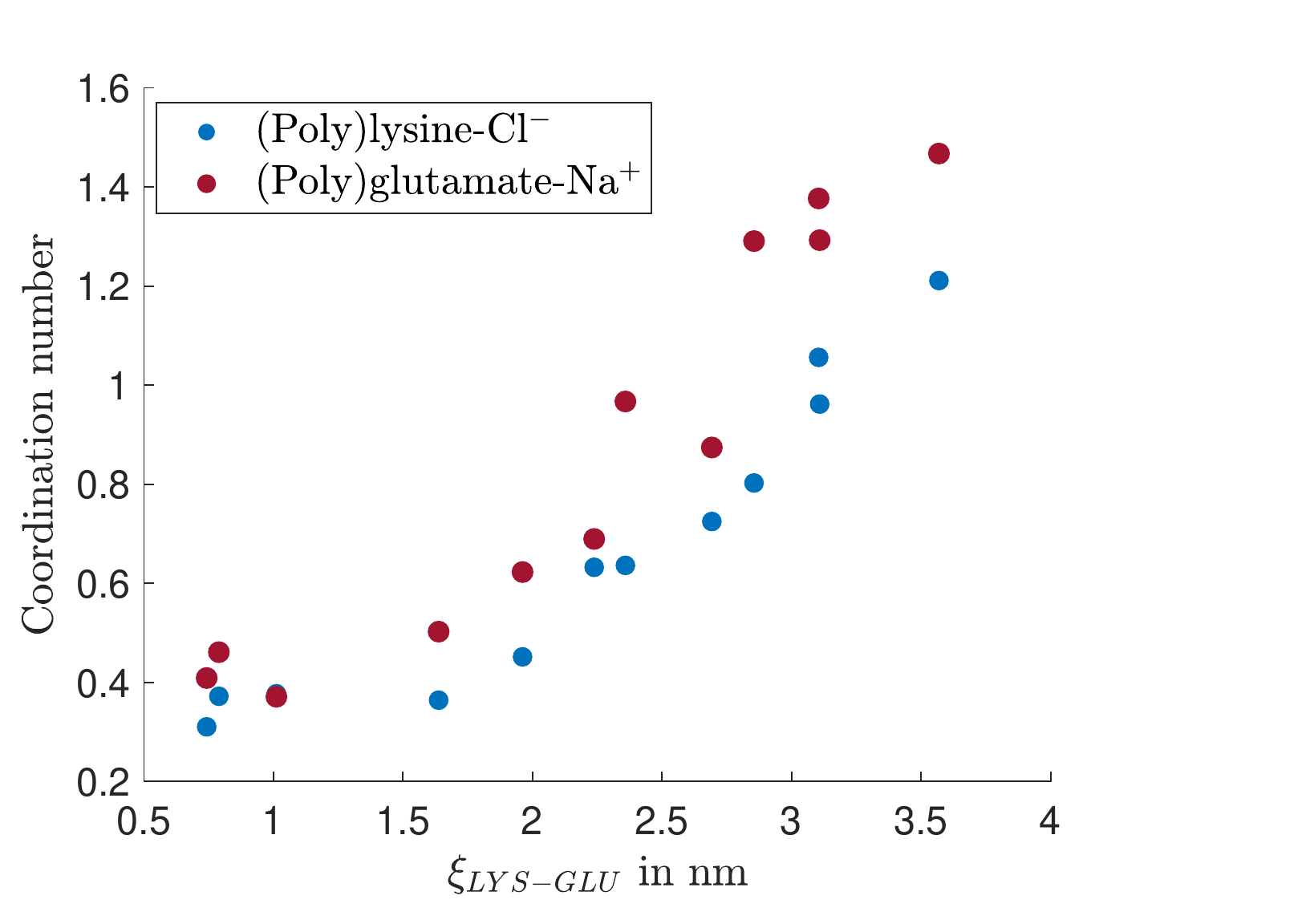}
    \caption{Coordination number, poly(lysine)-Cl$^-$ \& poly(glutamate)-Na$^+$}\label{fig:7c}
  \end{subfigure}
  \caption{Radial distribution function from TIP3P-AMBER with $C^{excess}_{NaCl}$=0M (a) between specific sites of poly(lysine) with the highest positive partial charges and Cl$^-$ counterions, and (b) between specific sites of poly(glutamte) with the highest negative partial charges and Cl$^-$ counterions.  The average number of counterions that reside in the first and second shell of the polypeptides at different distance of separation is shown in (c)}
  \label{fig:7}
\end{figure}

The counterion-polyion correlations can be quantified via the pair distribution function between counterions and the charged sites on the polyions.
Specific molecular sites of both polypeptides with the highest partial charge are categorized together and the radial distribution function between each of the these sites and the counterions are calculated. For poly(glutamate) these sites consist of the carbonyl atoms and for poly(lysine), these sites comprise the terminal nitrogen, hydrogen and carbon. These radial distribution functions are shown in \cref{fig:7a} and \subref{fig:7b}.  A substantial increase in the first and second peak of the radial distribution function can be observed for both poly(lysine)-Cl$^-$ and poly(glutamate)-Na$^+$ when they are not interacting with each other.  The magnitude of the first peak increases by a factor of 5 and 35, respectively, for 
poly(lysine)-Cl$^-$ and poly(glutamate)-Na$^+$.  

A coordination number calculation between the polypeptides and the counterions for the same system as above at different distance of separation between poly(lysine) and poly(glutamate) is also performed. This is done by defining a sphere of size 0.55 nm around both polypeptides, which approximately corresponds to the second shell (as seen in \cref{fig:7a}) and averaging the number of counterions found within that shell over all timesteps. Since a switch of driving factors for complexation from being entropically-driven to energetically-driven (as shown in \cref{fig:3e}) is observed at $\approx$ 1.5 nm for TIP3P-AMBER, it is important to investigate if this switch results from the counterions being bound to the polypeptides at that distance. If such is the case, it would be expected that there is a rapid increase in the coordination number of the ions at 1.5 nm. The coordination number plot is shown in ~\cref{fig:7c}, which indicates that there is no rapid increase in the coordination number at $\approx$ 1.5 nm. Rather, an almost a linear increase is seen as the polypeptides transition from a complexed to a more dilute region where they are not interacting with each other.

It is important to note that although there is a steep increase in the first and second peaks of the radial distribution function between the polypeptides and counterions as they are brought to a region of no interaction, the average number of ions that reside in the first and second shell of the polypeptides is still not substantially high. Even when the polyelectrolytes are not interacting with each other, only 12 - 16\% of the total counterions reside near them.   A movie of the trajectory of the system of polypeptides in the non-interacting regime is available at \href{https://figshare.com/articles/Oppositely-charged_Polypeptides_at_a_Distance_of_Separation_of_no_Interaction/9209489}{https://doi.org/10.6084/m9.figshare.9209489}, where it is evident that although the polypeptides have a large number of counterions in proximity (which can also be seen in \cref{fig:6b} and \subref{fig:6c}), the counterions are not bound or condensed, however, and freely move around the polypeptides and continuously leave and enter the coordination shells. Therefore the entropy of complexation does not arise from a loss of translational entropy of the counterions but rather from the increase in the probability of the counterions to be in close proximity with the polypeptides.

\section{Conclusion}

We study the driving force for the complexation of poly(lysine) and poly(glutamate) oligomers using molecular dynamics simulations of an atomistic and two coarse grained models.  Results from all three forcefields are in qualitative agreement for the potential of mean force and driving force for complexation.   The agreement between force fields, which have very different treatment for water, suggests that the solvent does not play a dominant role in the complexation process.   For peptides where every residue is charged, the driving force is entropic in all cases except when there are no small ions present, in which case it becomes energetic.  We conclude that the entropy of the counterions is the important physical reason for polyelectrolyte complexation.  When the charge density of the peptides is decreased, and there are correspondingly fewer counterions, the driving force becomes energetic.  This further supports the notion that the counterion entropy contributes to the driving force for complexation.

An interesting point is that all the systems studied are well over the counterion condensation threshold and the ratio $l_B/l \approx$1.8-3.2.
We do not see any dynamic localization of the counterions, which have non-zero self-diffusion constants and sample the entire volume independent of the separation of the polyions.  As a caveat, we mention that there are several differences between these chemically realistic models and the interpretation on the basis of phenomenology of simplified models might be problematic.  The peptides are quite short, and the charge moities are not on the backbone but rather on side-chains.  In addition, the non-electrostatic contributions to the interactions are likely to be significant.  Further study of the complexation of realistic polymers should help our understanding of the physics of the phase behavior of experimental systems.

\section{Author Information}
\textbf{Corresponding Author}\\
*E-mail: \href{yethiraj@chem.wisc.edu}{yethiraj@wisc.edu}\\
\textbf{ORCID}\\
Arun Yethiraj: \href{http://orcid.org/0000-0002-8579-449X}{0000-0002-8579-449X}\\
\textbf{Notes}\\
The authors declare no competing financial interest.
\begin{acknowledgement}

This research work was supported by the National Science Foundation through Grant No. CHE-1856595. We thank Kyeong-Jun Jeong and Dr. Ajay Muralidharan for stimulating and useful discussion regarding this project. All simulations presented here were performed in computational resources provided by UW-Madison Department of Chemistry HPC Cluster under NSF Grant No. CHE-0840494 and the UW-Madison Center for High Throughput Computing (CHTC).
\end{acknowledgement}

\bibliography{Coacervates}

\end{document}